\documentclass [12pt,a4paper] {article}
\usepackage[backend=biber]{biblatex}
\usepackage{amssymb,amsmath} 	
\author {a.author}
\usepackage{graphicx}
\usepackage{subcaption}
\usepackage{setspace}
\usepackage{array}
\usepackage{epsf, color, verbatim}
\usepackage{amssymb,amsmath}
\usepackage{soul}
\usepackage{color}
\usepackage{epsf, color, verbatim}
\usepackage{amssymb,amsmath}
\usepackage{soul}
\usepackage{color}
\usepackage[usenames,dvipsnames]{xcolor}
\usepackage{caption} 
\usepackage{booktabs}
\usepackage[ruled,vlined]{algorithm2e}
\usepackage{tabulary}
\usepackage{biblatex}

\newcommand{\bfmath}[1]{\mbox{\boldmath$#1$\unboldmath}}
\newtheorem{assumption}{Assumption}
\newenvironment{proof}[1][Proof]{\noindent\textbf{#1.} }{\ \rule{0.5em}{0.5em}}
\newtheorem{theorem}{Theorem}

\title{
\textbf{On Monitoring High-Dimensional Processes with Individual Observations}}
\small{\author{Mohsen Ebadi \\
 \makeatother
  {\small \it  University of Waterloo, Waterloo, ON N2L 3G1, Canada} \\
 Shoja'eddin Chenouri 
  \\{\small \it University of Waterloo, Waterloo, ON N2L 3G1, Canada} \\
Stefan H. Steiner\\  {\small \it  University of Waterloo, Waterloo, ON N2L 3G1, Canada}
}
\date{}
}
\begin {document}

\maketitle
\begin {abstract}
Modern data collecting methods and computation tools have made it possible to monitor high-dimensional processes. In this article, Phase II monitoring of high-dimensional processes is investigated when the available number of samples collected in Phase I is limitted in comparison to the number of variables. A new charting statistic for high-dimensional multivariate processes based on the diagonal elements of the underlying covariance matrix is introduced and a unified procedure for Phase I and II by employing a self-starting control chart is proposed.  To remedy the effect of outliers, we adopt a robust procedure for parameter estimation in Phase I and introduce the appropriate consistent estimators. The statistical performance of the proposed method is evaluated in Phase II  through average run length (ARL) criterion in the absence and presence of outliers and reveals that the proposed control chart scheme effectively detects various kinds of shifts in the process mean. Finally, we illustrate the applicability of our proposed method via a real-world example.
\end {abstract}
{\bf Keywords:} High-dimensional multivariate process; Phase II monitoring; Self starting control chart; Statistical process monitoring.
\section{Introduction}
Multivariate statistical process monitoring (MSPM) techniques have been extensively used to detect shifts in the parameters of multivariate processes. The well-known Hotelling's $T^{2}$ control chart is usually recommended for monitoring the mean of multivariate statistical process control with individual observations. Recently, modern data collecting and computation tools have made it possible to monitor high-dimensional processes. However, typical MSPM approaches used to monitor high-dimensional processes are frequently hampered by high-dimensional settings; the phenomenon also known as ``curse of dimensionality". This is mainly because the sample covariance matrix used in the methods based on $T^{2}$ statistic is singular. Despite many research papers being published on multivariate control charts to monitor the process mean (see for example Reynolds and Cho (2006), Reynolds and Stoumbos (2008), Woodall and Montgomery (2014) and Bersimis, et al. (2007) for discussions and reviews of multivariate control charts), monitoring changes in the mean vector for high dimensional multivariate processes has received little attention in the literature. We discuss the few exceptions in the following. Under the ``sparsity" assumption, which means that there are only a small set of variables responsible for the process change,
Wang and Jiang (2009) proposed using a forward selection algorithm combined with a Shewhart-type control chart, referred to as the VS-MSPC chart. Another variable selection (VS)-based approach, the adaptive absolute shrinkage and selection operator (LASSO), which identifies potentially altered variables, was proposed by Zou and Qiu (2009). Capizzi and Masarotto (2011) combined the least angle regression with MEWMA to monitor both the mean and variability. Jiang et al. (2012) investigate the impact of mean shifts estimation on the probability of accurately identifying changed variables and suggest a variable selection-based MEWMA (VS-MEWMA) control chart which is more sensitive to the small shifts in the mean vectors.  Abdella et al. (2017) used forward variable selection as a pre-diagnosis and it was integrated into the MCUSUM chart. Kim et al. (2020) proposed a penalised likelihood-based technique based on $L_2$ norm regularisation that shrinks all process mean estimates toward zero. Their proposed chart is efficient in monitoring high-dimensional processes since it has a closed-form solution as well as probability distributions of the monitoring statistic under null and alternative hypotheses. 

The aforementioned literature mainly focused on the problem of Phase II monitoring of high dimensional processes when there are enough data with no outlier observations in Phase I. 
However, in practical situations, the Phase I data are limitted and outliers may exist in Phase I data. For the situation where a reference dataset is not large enough to estimate the process parameters, self-starting methods that handle sequential monitoring by using the successive process readings to update the parameter estimates and simultaneously check for the out-of-control conditions exist in the literature. See, for example, Sullivan and Jones-Farmer (2002) and Hawkins and Maboudou-Tchao (2007). More recently, Chen et al. (2016) introduced a new nonparametric method for monitoring location parameters when only a small reference dataset is needed. Still, they assumed that $m>p$ and that the Phase I dataset is outlier free. However, In the high-dimensional situation, more robust estimators of the process parameters are required if some outliers contaminate the data. In many practical situations, the covariance matrix can not be meaningfully estimated from the original data due to the ``curse of dimensionality”. This research proposes a new self-starting control chart for Phase II monitoring of a high-dimensional process. In our new chart, rather than estimating all elements of the covariance matrix, we only estimate it's diagonal elements. By using a robust method to estimate the parameters, we show our approach is very effective in Phase II monitoring of the process mean, especially when the sample size is small compared to the number of variables.
The remainder of this paper is organized as follows. In Section 2, we develop our proposed charts based on the diagonal elements of the sample covariance matrix and a (unified) self-starting approach for Phase II monitoring is then proposed. In Section 3, the performance of the proposed chart is evaluated in terms of Average Run Length via Monte Carlo simulations. In Section 4, a real example is employed to show the applicability of the proposed methodology. Section 5 concludes our paper.

\section{Monitoring method for high-dimensional process}
This section proposes a new control chart based on the characteristics and limitations of high-dimensional processes. We then propose a self-starting approach for Phase II monitoring by using a robust estimation approach for the underlying parameters of the in-control processes when the historical data is limitted. 
\subsection{A new control chart}
Consider the problem of monitoring a multivariate process with $p$ quality characteristics $\mathbf{X}=(X_1,\,X_2,\,\dots,\,X_p)'$. It is assumed that there are $m$ independent and identically distributed (i.i.d) historical (reference) observations $\mathbf{X}_{1},\,\mathbf{X}_{2},\,\dots,\,\mathbf{X}_{m} $  collected for Phase I analysis. Under an in-control situation, we assume that the process follows a multivariate normal distribution with mean vector $\bfmath{\mu}$ and covariance matrix $\bfmath{\Sigma}$. For Phase II monitoring, the $i$th future observation, $\mathbf{X}_{i},\,i=m+1,\,m+2,\,\dots\,$ is collected to be monitored over time. A typical approach for monitoring the mean of such a process in Phase II uses the $T^{2}$ statistic based on the Mahalanobis distance (Mahalanobis 1936) as follows:
\begin{equation}\label{e1}
T^{2}_{i}=({\mathbf{X}_{i}}-\bfmath{\mu})'\bfmath{\Sigma}^{-1}({\mathbf{X}_{i}}-\bfmath{\mu}),\qquad i=m+1,\,m+2,\,\cdots\,.
\end{equation}
A Phase I control chart, however, can be obtained by replacing $\bfmath{\mu}$ and $\bfmath{\Sigma}$ with the sample mean and covariance matrix, respectively (Bersimis et al. 2007). A  large value of $T^{2}_{i}$ leads to rejection of the null hypothesis that the observation $\mathbf{X}_{i}$ follows $N_p(\bfmath{\mu},\bfmath{\Sigma})$ and consequently, if the value of the test statistic plots above the prespecified control limit, the chart signals an out-of-control situation. When the in-control parameters of the process are known or estimated at the end of Phase I, the $T^{2}_{i}$ statistic follows a $\chi^{2}$-distribution with $p$ degrees of freedom. This is called a Phase II $\chi^{2}$ control chart and has the upper control limit of $\chi^{2}_{_{1-\alpha,\,p}}$ and the lower control limit of zero (see, for example, Bersimis et al. 2007). The conventional estimator of the parameters $\bfmath{\mu}$ and $\bfmath{\Sigma}$ in Phase I are the sample mean vector and the sample covariance matrix, respectively. But when $p>m$, the standard sample covariance matrix is singular and cannot be inverted, so the $T^{2}$ statistic becomes ill-defined. In practice, the available number of Phase I samples is usually limited, and users of process monitoring approaches do not want to wait until many samples accumulate. In this paper, we propose using the diagonal matrix consisting of only the diagonal elements of the sample covariance matrix, obtained from the Phase I analysis, to replace the covariance matrix in calculating the critical distance for multivariate process monitoring in Phase II. Note that the diagonal elements of a covariance matrix can be estimated by as few as two individual observations, so it will not be affected by high dimensionality. Consider the sampling epoch $i$ and denote the individual observation of the $j$th quality characteristic variable by $X_{i\,j},$ where $i=m+1,\,\dots\,$, and $j=1,\,\dots,\,p$. Let $\bfmath{\sigma}=({\sigma}_{_{1\,1}},\,\dots,\,{\sigma}_{_{p\,p}})'$ denote the vector of in-control variances of the $p$ variables obtained from the diagonal elements of $\bfmath{\Sigma}$. If we define $\bfmath{D}=\rm{diag}({\sigma}_{_{1\,1}},\,\dots,\,{\sigma}_{_{p\,p}})$, then the corresponding modified Mahalanobis distance is:
\begin{equation}\label{e2}
M^{2}_{i}=M^{2}_{i}(\bfmath{\mu},\mathbf{D})=({\mathbf{X}_{i}}-\bfmath{\mu})'\mathbf{D^{-1}}({\mathbf{X}_{i}}-\bfmath{\mu})=\sum_{j=1}^{p}\frac{(X_{ij}-{\mu}_{_j})^{2}}{\sigma_{_{j\,j}}}\,,
\end{equation}
where ${\mu}_{_{j}}$ denotes the $j$th element of the vector $\bfmath{\mu}$. Let $\lambda_{1},\,\lambda_{2}, ..., \,\lambda_{p}$ denote the eigenvalues of the in-control correlation matrix  $\bfmath{\rho}= \mathbf{D}^{-\frac{1}{2}}\,\bfmath{\Sigma}\,\mathbf{D}^{-\frac{1}{2}}$ and $\bfmath{\Lambda}={\rm diag}(\lambda_{1},\,\lambda_{2},\,\dots, \,\lambda_{p})$, then by the eigenvalue decomposition $\bfmath{\rho}=\bfmath{\Gamma}\,\bfmath{\Lambda}\,\bfmath{\Gamma}'$, where columns of the orthogonal matrix $\bfmath{\Gamma}$ constitute an orthogonal basis of eigenvectors of $\bfmath{\rho}$. Now using the transformation $\mathbf{Z}_{_i}=\bfmath{\rho}^{-\frac{1}{2}}\,\mathbf{D}^{-\frac{1}{2}}({\mathbf{X}_{i}}-\bfmath{\mu})$, 
\begin{align*}
 M^{2}_{i}&= ({\mathbf{X}_{i}}-\bfmath{\mu})'\,\mathbf{D^{-1}}({\mathbf{X}_{i}}-\bfmath{\mu})\\
 &=\mathbf{Z}_{_i}'\bfmath{\rho}\,\mathbf{Z}_{_i}=\mathbf{Z}_{_i}'\,\bfmath{\Gamma}'\,\bfmath{\Lambda}\,\bfmath{\Gamma}\,\mathbf{Z}_{_i}\\
 &=\bfmath{\xi}_{_i}'\,\bfmath{\Lambda}\,\bfmath{\xi}_{_i}\,,
\end{align*}
where $\bfmath{\xi}_{_i}=(\xi_{_{i1}},\,\ldots ,\,\xi_{_{ip}})'=\bfmath{\Gamma}\,\mathbf{Z}_{i}$. This shows that the modified distance 
$M^{2}_{i}$ can be rewritten as $M^{2}_{i}=\sum_{j=1}^{p}\lambda_{j}\xi_{_{ij}}^{2},$ where $\xi_{_{ij}}$, for   $j=1,\,\ldots,\,p$, are i.i.d standard normal random variables. Since $\xi_{_{ij}}^{2}\sim\chi^{2}_{_{(1)}}$, the distance $M^{2}_{i}$ is the weighted sum of i.i.d. random variables with $\chi^{2}_{_{(1)}}$ distribution, and thus the mean and variance of $M^{2}_{i}$ are given by
\begin{equation}\label{e3}
{\rm E}(M^{2}_{i})={\rm E}\left(\sum_{j=1}^{p}\lambda_{j}\xi_{_{ij}}^{2}\right)=\sum_{j=1}^{p}\lambda_{j}{\rm E}(\xi_{_{ij}}^{2})=\sum_{j=1}^{p}\lambda_{j}={\rm tr}(\bfmath{\rho})=p
\end{equation}
\begin{equation}\label{e4}
{\rm Var}\left(M^{2}_{i}\right)=\rm{Var}\left(\sum_{j=1}^{p}\lambda_{j}\xi_{_{ij}}^{2}\right)=\sum_{j=1}^{p}\lambda_{j}^{2}\rm{Var}(\xi_{_{ij}}^{2})=2\sum_{j=1}^{p}\lambda_{j}^{2}=2\,\textrm{tr}(\bfmath{\rho}^2),
\end{equation}
where $\textrm{tr}(\bfmath{A})$ represents the trace of matrix $\bfmath{A}$. Using the mean and variance of the modified distance, one can define the statistic:
\begin{equation}\label{e5}
U_{i}=\frac{M^{2}_{i}(\bfmath{\mu},\mathbf{D})-p}{\sqrt{2\,{\rm tr}(\bfmath{\rho}^2)}}\,.
\end{equation}
To derive our asymptotic results, we make the following assumptions: 

\begin{assumption}\label{A1} 
For $i=1,\,\dots,\,6\,$, we assume that  $0<\lim\limits_{p\rightarrow{\infty}}\,p^{-1}\,{\rm tr}\left(\bfmath{\rho}^i\right)<\infty\,$.
\end{assumption}

\begin{assumption}\label{A2}
The eigenvalues $\lambda_i$ of the correlation matrix $\bfmath{\rho}$ satisfy $\lim\limits_{p\rightarrow{\infty}} \underset{1\leq i\leq p}{\max} \,\lambda_i\, p^{-1/2}=0\,$.
\end{assumption}

\begin{assumption}\label{A3}
The dimension $p$ grows with sample size $m$ at a rate of $p=O(m^{1/\zeta})$ with $1/2<\zeta\leq{1}\,$.

\end{assumption}

\begin{assumption}\label{A4}
For some $0 <\gamma <\zeta/2$, $\lim\limits_{p\rightarrow{\infty}} \underset{1\leq i\leq p}{\max}\,\lambda_i\,p^{-\gamma}<\infty\,$.
\end{assumption}
 Since Assumptions 1-2 imply $\max\limits_{1\le i\le p}\lambda_i^2/\sum\limits_{j=1}^p\lambda_j^2=o(1)$, with a direct application of the H\'{a}jek–\v{S}id\'{a}k central limit theorem (c.f. DasGupta 2008) we can show that for any given $i=1,\,\dots,\,m$ the statistic $U_{i}$ has an asymptotic $N(0,1)$ distribution as $p\rightarrow\infty$. It is worth mentioning that the assumptions provided above are not very restrictive. For example, Assumption 1 implies that the growth rate for traces of powers of the correlation matrix should not be higher than $p$. Thus, it can be valid even when some strong pairwise correlations exist among data like popular cases of autoregressive (AR) or moving average (MA) structures. For example, we can easily show that ${\rm tr}(\rho^2)=\sum\limits_{i=1}^p\sum\limits_{j=1}^p \rho_{ij}^2$. Thus, for an AR correlation structure we have ${\rm tr}(\rho^2)=p+2\sum\limits_{k=1}^p\rho^{2k}=O(p)$. This is also true for some MA or banded correlation structures, i.e. ${\rm tr}(\rho^2)=p+2\sum\limits_{i=1}^p\rho_{i,\,i+1}^2=O(p)$, satisfying Assumption 1. Inspired by the work of Srivastava and Du (2008) on the one-sample test of the mean vector in a high-dimensional setting, Ro et al. (2015) used the aforementioned asymptotic result for $U_{i}$ in $p$, for outlier detection purposes in high-dimensional datasets. From simulation studies, we observed that the asymptotic normality of $U_{i}$ in $p$ fails to accurately approximate the tails of the distribution $U_{i}$. Besides, investigating Table 1 of Ro et al. (2015) reveals the estimated probability of false positives are overestimated for moderate dimensions and small nominal Type I errors $\alpha=0.01,\, 0.05$. Since small values of $\alpha$ such as 0.005 are common in constructing control charts, we need a good approximation of the quantiles of the exact distribution of $U_{i}$. 

To improve the accuracy of approximations, we employ the Cornish–Fisher asymptotic expansion of quantiles, which uses higher-order moments of $U_{i}$ to account for the effect of non-normality. These expansions were first developed by Cornish and Fisher (1938) and Fisher and Cornish (1960). The expansion can be derived by inverting the Edgeworth expansion of the distribution of $U_{i}$. See Hall (1983), Small (2010), and Polansky (2011). 

\begin{theorem} 
Using the second-order Cornish-Fisher expansion, the upper percentile of statistic $U_{i}$ at significance level $\alpha$ is given by 
\begin{align}\label{CFexp}
\omega_{\alpha,\,p}= z_{\alpha}+\frac{4\,\textrm{tr}(\bfmath{\rho}^{3})\,(z_{\alpha}^2-1)}{3\left[2\,{\rm tr}(\bfmath{\rho}^{2})\right]^\frac{3}{2}}+\frac{\textrm{tr}(\bfmath{\rho}^4)}{2\left[{\rm tr}(\bfmath{\rho}^2)\right]^{2}}\,(z_{\alpha}^3-3z_{\alpha})
+\frac{2\left[{\rm tr}(\bfmath{\rho}^{3})\right]^2}{9\left[{\rm tr}(\bfmath{\rho}^{2})\right]^{3}}\,({5z_{\alpha}-2z_{\alpha}^3})\,,
\end{align}
where $z_\alpha$ is the upper $100\alpha\%$ percentile of the standard normal distribution. 
\end{theorem} 
A proof of the formula \eqref{CFexp} is provided in Appendix A. Depending on how small the type I error is set, we can either use a first or second order Cornish-Fisher expansion. Although we have given the second-order in \eqref{CFexp}, in our simulation study in Section 3, we only use the first-order expansion of the Cornish-Fisher
\begin{align}\label{CFexp1}
\omega_{\alpha,\,p}\approx z_{\alpha}+\frac{4\,{\rm tr}(\bfmath{\rho}^{3})\,(z_{\alpha}^2-1)}{3\left[2\,{\rm tr}(\bfmath{\rho}^{2})\right]^\frac{3}{2}}\,,
\end{align}
as it suffices to achieve good results.

The discussion above suggests a control chart for monitoring $\mathbf{X}_i$ based on its respective $U_i$ value when $\bfmath{\Sigma}$ or equivalently $\bfmath{\rho}$ is known or properly estimated.  The proposed control chart triggers an out-of-control alarm whenever 
\begin{equation}\label{chart}
Z_{i}=U_{i}-\frac{4\,\textrm{tr}(\bfmath{\rho}^{3})\,(z_{\alpha}^2-1)}{3\left[2\,{\rm tr}(\bfmath{\rho}^{2})\right]^\frac{3}{2}}>z_{\alpha},
\end{equation}


When the process is in-control, a new observation follows $N_{p}(\bfmath{\mu},\bfmath{\Sigma})$, while in the out-of-control situation the observations follow $N_{p}(\bfmath{\mu}_1,\bfmath{\Sigma})$.
Consequently, assuming $\bfmath{\mu}_1-\bfmath{\mu}=\bfmath{\delta}$, the asymptotic Type II error probability of the proposed control chart as $p\rightarrow\infty$ can be derived. Since for the Cornish-Fisher expansion $\omega_{\alpha,\,p}\rightarrow  z_{\alpha}$ as $p\rightarrow \infty$, the following lemma provides an asymptotic type II error probability.

\begin{theorem}\label{thm2}
Under Assumptions 1-2, and for $\bfmath{\delta}=O(p^{-1-\epsilon/2})$, $\epsilon>0$, we have
\begin{equation*}\lim\limits_{p\rightarrow \infty}\left[\Pr\left(U_i\le z_\alpha\,\mid \bfmath{\mu}_1\right)-\Phi\left(z_\alpha
-\frac{{\bfmath{\delta}}^{\prime}\mathbf{D}^{-1}\bfmath{\delta}}{\sqrt{2\,{\rm tr}(\bfmath{\rho}^{2})}}\right) \right]=0
\end{equation*}
\end{theorem}
\begin{proof} First notice that 
\begin{equation*}
M^{2}_{i}(\bfmath{\mu}_1,\mathbf{D})
=M^{2}_{i}(\bfmath{\mu},\mathbf{D})+\bfmath{\delta}^\prime\,\mathbf{D}^{-1}\,\bfmath{\delta}-2\,\bfmath{\delta}^\prime\,\mathbf{D}^{-1}\,(\mathbf{X}_{i}-\bfmath{\mu})\,.
\end{equation*}
Under $\mathbf{X}_i\sim N_{p}(\bfmath{\mu}_1,\bfmath{\Sigma})$, we have ${\rm E}\left[\bfmath{\delta}'\,\mathbf{D}^{-1}\,(\mathbf{X}_{i}-\bfmath{\mu}) \right]=\bfmath{\delta}^\prime\,\mathbf{D}^{-1}\,\bfmath{\delta}$ and 
\begin{align*}
{\rm Var}\left[\bfmath{\delta}'\,\mathbf{D}^{-1}\,(\mathbf{X}_{i}-\bfmath{\mu}) \right]&={\rm Var}\left[\bfmath{\delta}'\,\mathbf{D}^{-1}\,(\mathbf{X}_{i}-\bfmath{\mu}_{1}) \right]\\
&={\rm Var}\left[\sum\limits_{j=1}^{p}\delta_j\frac{(X_{ij}-{\mu}_{_{1\,j}})}{\sigma_{_{j\,j}}}\right]\\
&={\rm Var}\left[\sum\limits_{j=1}^{p}\frac{\delta_j}{\sqrt{\sigma_{_{j\,j}}}}\cdot\frac{(X_{ij}-{\mu}_{_{1\,j}})}{\sqrt{\sigma_{_{j\,j}}}} \right]\\
&=\sum\limits_{j=1}^{p}\frac{\delta_j^2}{\sigma_{_{j\,j}}}+\sum\limits_{j\neq k}^{p}\frac{\delta_j\,\delta_k}{\sqrt{\sigma_{_{j\,j}}\,\sigma_{_{k\,k}}}}\cdot\rho_{_{j\,k}}\\
&=O(p^{-1-\epsilon})+O(p^{-\epsilon})=O(p^{-\epsilon})
\end{align*}
that is ${\rm Var}\left[\bfmath{\delta}'\,\mathbf{D}^{-1}\,(\mathbf{X}_{i}-\bfmath{\mu}) \right]\rightarrow 0$ as $p\rightarrow \infty$. Thus $\bfmath{\delta}'\,\mathbf{D}^{-1}\,(\mathbf{X}_{i}-\bfmath{\mu})\overset{p}{\rightarrow}\bfmath{\delta}^\prime\,\mathbf{D}^{-1}\,\bfmath{\delta}$, and $M^{2}_{i}(\bfmath{\mu}_{1},\mathbf{D})
\overset{p}{=}M^{2}_{i}(\bfmath{\mu},\mathbf{D})-\bfmath{\delta}^\prime\,\mathbf{D}^{-1}\,\bfmath{\delta} $.   
Now under $\bfmath{\delta}=O(p^{-1-\epsilon/2})$, $\epsilon>0$ and Assumptions 1 and 2,
\begin{align*}
 \lim\limits_{p\rightarrow \infty}\Pr(U_i\le z_\alpha\,\mid \bfmath{\mu}_1)&=
  \lim\limits_{p\rightarrow\infty}
 \Pr\left(\frac{M^{2}_{i}(\bfmath{\mu}_{1},\mathbf{D})+\bfmath{\delta}^{\prime}\mathbf{D}^{-1}\bfmath{\delta}-p}{\sqrt{2\,{\rm tr}(\bfmath{\rho}^{2})}}\le z_\alpha\,\mid \bfmath{\mu}_1\right) \\ 
 &= \lim\limits_{p\rightarrow\infty}
 \Phi\left(z_\alpha
-\frac{{\bfmath{\delta}}^{\prime}\mathbf{D}^{-1}\bfmath{\delta}}{\sqrt{2\,{\rm tr}(\bfmath{\rho}^{2})}}\right) \,. 
\end{align*}
\end{proof}

As the proposed chart is a Shewhart-type chart, the in-control and out-of-control ARLs of the proposed control chart are
\begin{equation}\label{arls}
{\rm ARL}_{_0}=\frac{1}{1-\alpha}\quad \text{ and } \quad{\rm ARL}_{_1}=\frac{1}{1-\beta}\,,
\end{equation} 
where $\beta$ represents the probability of type II error when the process is out-of-control and can be asymptotically calculated by using Theorem 2. 

In order to use \eqref{chart} in Phase II, proper estimates of the parameters $\bfmath{\mu}$,  $\mathbf{D}$, ${\rm tr}(\bfmath{\rho}^{2})$, and ${\rm tr}(\bfmath{\rho}^{3})$ in Phase I are needed to obtain good results in Phase II. 
For estimating $\mathbf{D}$, ${\rm tr}(\bfmath{\rho}^{2})$ based on $m$ observations in Phase I, one can use the suggested estimator by Srivastava and Du (2008) where used a consistent estimator under Assumption 1 and 3 as follows 
$$\frac{1}{p}\left[{\rm tr}(\bfmath{R}^{2})-\frac{p^2}{m}\right]-\frac{1}{p}{\rm tr}(\bfmath{\rho}^{2})\rightarrow 0 \quad \text{as } n, \,p \rightarrow\infty\,.$$

where the sample correlation matrix $\bfmath{R}$ in Phase I can be given from
\begin{equation}\label{e6} 
 \bfmath{R}= \bfmath{D}_{_S}^{-\frac{1}{2}}\,\bfmath{S}\,\bfmath{D}_{_S}^{-\frac{1}{2}}\,,
\end{equation}
where $\bfmath{S}$ is the sample covariance matrix and $\bfmath{D}_{_S}$ denotes the diagonal matrix of the sample variances in $\bfmath{S}$. Besides, the following consistent estimator of ${\rm tr}(\bfmath{\rho}^{3})$ can be used as $(m,p)\rightarrow\infty$. 
\begin{equation}\label{e8}
{\rm tr}(\mathbf{R}^3)-\frac{3p}{m}\,{\rm tr}(\mathbf{R}^2)+\frac{2\,p^3}{m^2}
\end{equation}

A proof is provided in Ebadi et al. 2021.

In order to remedy the effects of outliers in Phase I, we apply a methodology for the robust estimation of the parameters proposed by Ebadi et al. (2021) through modifying re-weighted minimum diagonal product (RMDP) algorithm with Cornish-Fisher expansion. They also proposed a finite sample correction coefficient for better convergence via a simulation study and careful numerical evaluations defined as follows
\begin{equation}\label{e11}
c_{_{p,\,m}}=1+\frac{2\,p}{m\sqrt{{\rm tr}(\bfmath{R}^{2})-\frac{p^2}{m}}}
\end{equation} 
which under Assumptions 1 and 3, $c_{_{p,\,m}}\overset{p}\longrightarrow 1$ as $m,\,p\rightarrow \infty$.  

We denote the estimated parameters from RMDP algorithm proposed by Ebadi et al. (2021) as $\widetilde{\bfmath{\mu}}$, $\widetilde{\bfmath{D}}$, $\widehat{{\rm tr}\,(\bfmath{\rho}^2)}_{_{\rm RMDP}}$, and $\widehat{{\rm tr}\,(\bfmath{\rho}^3)}_{_{\rm RMDP}}$ and will use these estimates as initial estimates in our self starting control chart. These estimates will be then updated as new observations will appear. In the next subsection, we introduce a self-starting procedure for monitoring high-dimensional data.


\subsection{A self-starting procedure for Phase II}
In this section, we provide a procedure to perform Phase I analysis and Phase II monitoring subsequently. The steps of the proposed procedure are the following:
\begin{itemize}
\item[i. ] Collect a historical sample of size $m$ from the multivariate  $N_{p}(\bfmath{\mu},\bfmath{\Sigma})$ for Phase I analysis.
\item[ii. ] Implement the robust procedure of Ebadi et al. (2021) (mentioned in Section 2.1) to derive the robust estimates $\widetilde{\bfmath{\mu}}$, $\widetilde{\bfmath{D}}$, $\widehat{{\rm tr}(\bfmath{\rho}^2)}_{_{\rm RMDP}}$, $\widehat{{\rm tr}(\bfmath{\rho}^3)}_{_{\rm RMDP}}$ and the finite sample correction factors. Then identify the potential outlying observations in Phase I data.  
\item[iii. ] Having obtained the estimates in step (ii), for a new observation $i=m+1$ in Phase II, use the control chart \eqref{chart}, presented in Section 2.1.
\item[iv. ] If the new observation is identified as an in-control observation, update the estimates in step (ii) by adding the new observation to the Phase I data.
\item[v. ] Repeat the steps (iii) and (iv) for the new observations $i=m+2,\, \dots$ until the control chart triggers an out-of-control alarm. 
\end{itemize}
Notice that steps (iii)-(v) define a self-starting Phase II control chart. As articulated in the paper Maboudou-Tchao and Hawkins (2011), to implement a self-starting control chart, one needs to have enough historical samples to obtain initial estimates of the process parameters. This makes the proposed Phase II chart less sensitive to the initial Phase I sample, and the nominal $ARL_0$ can be achieved more stably for different samples. In other words, in self-starting control charts, the process parameters are updated continually over the time of sampling, and also, the out-of-control condition is checked concurrently. As the in-control period increases, the estimated mean vector and covariance matrix converge to the true mean vector
and covariance matrix so that the asymptotic normality of the underlying statistic is achieved. There is an important difference between our proposed self-starting chart in step (iii)-(v) and those in the literature such as in Maboudou-Tchao and Hawkins (2011). Methods in the literature typically require at least $p+1$ initial process reading vectors to set up the initial non-degenerate estimates of parameters, while our proposed chart does not have this limitation since the estimation of the covariance matrix is not required.

Generally, as new observations are collected in Phase II, updating the covariance matrix estimate becomes challenging in high-dimensional cases. For this purpose, the method proposed by Quesenberry (1997) can be useful. Let
\begin{equation}\label{Q1}
\overline{\mathbf{X}}_{j-1}=\frac{1}{j-1}\,\sum_{i=1}^{j-1}\,\mathbf{X}_{i}\,, \quad
\mathbf{A}_j=\mathbf{X}_j-\overline{\mathbf{X}}_{j-1}\,, \quad
\mathbf{Q}_j=\mathbf{Q}_{j-1}+\frac{j-1}{j}\, \mathbf{A}_j\,\mathbf{A}_j^{T}\,.
\end{equation}
Quesenberry (1997) suggested the following updating formulas 
\begin{equation}\label{Q2}
\overline{\mathbf{X}}_{j}=\frac{1}{j}\left[\,(j-1)\,\overline{\mathbf{X}}_{j-1}+\mathbf{X}_{j}\,\right],\qquad
\mathbf{S}_j=\frac{1}{j-1}\,\mathbf{Q}_j
\end{equation}
to reduce the computational cost in calculating the sample mean and covariance matrix. See also Sullivan and Jones-Farmer (2002).
In the next section, we examine the performance of our proposed methods through simulation.

\section{Simulation Study}
This section investigates the performance of proposed methods through a simulation study both in the absence and presence of contaminated data. R Software is employed for this purpose with two scenarios, denoted by Scenario I and Scenario II, where in both scenarios the common mean vector is $\mathbf{0}_{p}$, while their covariance matrices are $\mathbf{I}_p$, and $\sigma_{ij}=(0.5)^{|i-j|}\quad {\rm for}\quad i,j=1,\dots,p$, respectively which are related to independent and autoregressive correlation structures. 

We evaluate the performance of the proposed chart in Phase II via the ARL criteria. Recall that ARL is the average number of samples taken until an out-of-control signal is observed and ${\rm ARL}_{_0}$ and ${\rm ARL}_{_1}$ are for in-control and out-of-control situations, respectively.  Large ${\rm ARL}_{_0}$ and small ${\rm ARL}_{1}$ are desirable to guarantee few false alarms and fast detection of process changes, respectively. In this paper, we use a total of 10,000 replications to estimate the ARL values. Any deviation of the CDF of $U_i$ from the standard normal leads to a very different ${\rm ARL}_{_0}$ of its corresponding control chart from the nominal values of ${\rm ARL}_{_0}$, which increases the false alarm rate of that control chart. Table \ref{table:arl0} presents the results of ${\rm ARL}_{_0}$ with and without using the Cornish-Fisher expansion when we use the exact parameters in Phase II for Scenarios 1 and 2. The UCLs for both charts are determined to give the nominal ${\rm ARL}_{_0}$. On the other hand, Table \ref{table:arl1} compares the results of ${\rm ARL}_{_1}$ with and without using the Cornish-Fisher expansion for different values of $p$ and $\alpha$ when the mean vector for  $20\%$ of observations has shifted by the vector $\mathbf{1}=(1,\,1,\,\dots,\,1)'$. In both Tables \ref{table:arl0} and \ref{table:arl1}, we also provided the nominal ${\rm ARL}_{_0}$ and ${\rm ARL}_{_1}$ using \eqref{arls}. It is clear from Tables \ref{table:arl0} and \ref{table:arl1}, when using the Cornish-Fisher expansion, both the estimated in-control and out-of-control ${\rm ARL}$s are generally much closer to their nominal values, while the original statistic $U_i$ proposed by Srivastava and Du (2008) gives very different ${\rm ARL}$s from their respective nominal values. Additionally, we performed other simulations with different values for $\bfmath{\Sigma}$, $\alpha$, and $p$. These simulations, not reported here, are consistent with the conclusions presented in Tables \ref{table:arl0} and \ref{table:arl1}. As expected,  the average in-control ARL values become closer to the nominal values by increasing the Phase I sample size $m$ as incorporating new in-control observations of Phase II into the estimation improves the estimates' accuracy.  Another important conclusion is that when $\alpha$ is smaller, the in-control ARL is generally much closer to the nominal value because more in-control observations will be involved in the estimation. Hence, we recommend using larger $m$ with smaller $\alpha$. Note that the initial samples in Phase I may affect the performance of the chart in Phase II and give different ${\rm ARL}_{_0}$s, but our simulation showed that having Phase I data with a sample size of about 200-300 is appropriate. Since our method only needs to estimate $p$ diagonal elements of the covariance matrix, obtaining accurate estimates are much affordable than methods based on $T^2$ which needs to estimate $p(p+1)/2$ elements. 

We also conduct a simulation to evaluate the effect of correlation on the proposed method. Note that high correlations may happen in applications when dimensionality $p$ is small, but for high dimensional cases it seems unlikely. For example as stated by Ahmadi-Javid and Ebadi (2021), in high-dimensional multiple stream processes (MSPs) as a particular multivariate processes with two sources
of variation, the streams are usually
independent or weakly correlated. As a sensitivity analysis, Table \ref{table:cor} investigates the effect of correlation on the calculated ${\rm ARL}_{_0}$ when  covariance matrix $\sigma_{ij}=(a)^{|i-j|}\quad {\rm for}\quad i,j=1,\dots,p$ is used and gradually change the value of $a$ from 0 to 0.9. We set the mean vector $\bfmath{\mu}$ to zero. In calculating ${\rm ARL}_{_0}$, we use the true values of the parameters. It can be observed from Table \ref{table:cor} that as the correlation between variables increases, the value of ${\rm ARL}_{_0}$ generally remains close to the nominal values for either $\alpha=0.01$ or 0.005. However, for a few cases with high correlation such as $a=0.9$ and small values of $p$, the simulated ${\rm ARL}_{_0}$ is slightly greater than the nominal value which is a positive point when the process is under control, but ${\rm ARL}_1$ may increases. We do not provide a study of the change in ${\rm ARL}_1$ by increasing $\rho$ to save space.  Typically, for the smaller values of ${\rm ARL}_1$, the change in this quantity is negligible in comparison to ${\rm ARL}_0$ values. However, since $M_{_i}(\bfmath{\mu},\,\mathbf{D})$ is a weighted sum of independent $\chi^2_{(1)}$ random variables, this shortcoming can be overcomed  by adopting the  Welch-Satterthwaite (W-S) $\chi^2$-approximation. More details are available in Satterthwaite (1941, 1946), Welch (1947), and Zhang et al. (2020) who recently showed the effectiveness of the Welch-Satterthwaite approximation in their proposed high-dimensional two-sample test statistic when the variables are highly correlated. Another solution for this phenomena, which maybe worth considering as future work, is to determine a finite sample correction coefficient for the case of highly-correlated multivariate process by using of an extensive simulation such that the control chart can achieve the nominal ${\rm ARL}_{_0}$.

\begin{table}[h!]

\scriptsize
\centering
\caption{Simulated ${\rm ARL}_{_0}$ for different values of $p$ and $\alpha$ when actual process parameters are used.}
\label{table:arl0}
\begin{tabular}{ |p{0.70cm}|p{0.6cm}|p{0.7cm}|p{0.6cm}|p{0.5cm}|p{0.6cm}|p{0.5cm}||p{0.6cm}|p{0.5cm}|p{0.6cm}|p{0.5cm}|p{0.6cm}|p{0.5cm}|  }
 \hline
\multicolumn{7} {|c|} {\bf Scenario 1} &\multicolumn{6} {|c|} {\bf Scenario 2}\\ \hline
&\multicolumn{2}{|c|}{$\alpha$=0.01} & \multicolumn{2}{|c|}{$\alpha$=0.005} & \multicolumn{2}{|c|}{$\alpha$=0.0027} &\multicolumn{2}{|c|}{$\alpha$=0.01} & \multicolumn{2}{|c|}{$\alpha$=0.005} & \multicolumn{2}{|c|}{$\alpha$=0.0027}\\ 
&\multicolumn{2}{|c|}{(${\rm ARL}_{_0}=100$)} & \multicolumn{2}{|c|}{(${\rm ARL}_{_0}=200$)} & \multicolumn{2}{|c|}{(${\rm ARL}_{_0}=370$)} &\multicolumn{2}{|c|}{(${\rm ARL}_{_0}=100$)} & \multicolumn{2}{|c|}{(${\rm ARL}_{_0}=200$)} & \multicolumn{2}{|c|}{(${\rm ARL}_{_0}=370$)}\\ 
\hline 
&  {\centering{\tiny With}} & {\centering{\tiny Without}}& {\tiny With} & {\tiny Without}& {\tiny With} & {\tiny Without}& {\tiny With} & {\tiny Without}& {\tiny With} & {\tiny Without}& {\tiny With} & {\tiny Without}\\
 &  {\centering{\tiny CF}} & {\centering{\tiny CF}}& {\tiny CF} & {\tiny CF}& {\tiny CF} & {\tiny CF}& {\tiny CF} & {\tiny CF}& {\tiny CF} & {\tiny CF}& {\tiny CF} & {\tiny CF}\\\hline
$p$=10& 104.8&\centering{39.3} &207.4&\centering{55.4}&376.3& \centering{77.7}&110.7 &\centering{32.5}&202.7&\centering{42.7}&356.5&55.1\\ \hline 
$p$=20&102.5 &\centering{45.5} &204.2 &\centering{70.3}&366.5 &\centering{100.3}&104.8 &\centering{36.3} &201.2&\centering{51.1}&347.7&67.5\\ \hline 
$p$=30&101.1&\centering{50.8}  &199.4 &\centering{79.9}&373.7 &\centering{119.1}&102.3 &\centering{39.9} &193.1&\centering{57.3}&346.1&77.9\\ \hline 
$p$=50&100.5 &\centering{56.8} &202.1 &\centering{92.6}&365.6&\centering{140.4} &100.4 &\centering{43.9}&196.9&\centering{67.4}&353.3&94.6\\ \hline 
$p$=80&100.4 &62 &202.2 &104.8&360.2 &165.3&102.2 &50.1 &193.8&76.7&354.9&113.8\\ \hline 
$p$=100&101.1 &63.9 &199.8 &109.1&373.2 &173.4&98.6 &52.1 &195.2&82.2&352.4&123\\ \hline 
$p$=150&101.7 &68.9 &197.7 &120.3&373.4&194.7 &99.1&56.3 &198.5 &94.2&362.8&142.1\\ \hline 
$p$=200&99.7 &71.1 &196.3 &127.3&370.1 &209.3&99.5 &60 &198.3&101.1&361.2&152\\ \hline 
\end{tabular}
\end{table}

\begin{table}[h!]

\tiny
\centering
\caption{Simulated ${\rm ARL}_{_1}$ for different values of $p$ and $\alpha$ when actual process parameters are used and $20\%$ of variables is shifted}
\label{table:arl1}
\begin{tabular}{ p{0.70cm}|p{1cm}|p{0.5cm}|p{0.75cm}|p{0.5cm}|p{0.55cm}|p{0.5cm}|p{0.55cm}|p{0.5cm}|p{0.55cm}|p{0.5cm}|p{0.55cm}|p{0.5cm}|p{0.5cm}  }
 \hline
\multicolumn{8} {c|} {\bf Scenario 1} &\multicolumn{6} {|c} {\bf Scenario 2}\\ \hline
\multicolumn{2} {c|} {}&\multicolumn{2}{|c|}{$\alpha$=0.01} & \multicolumn{2}{|c|}{$\alpha$=0.005} & \multicolumn{2}{|c|}{$\alpha$=0.0027} &\multicolumn{2}{|c|}{$\alpha$=0.01} & \multicolumn{2}{|c|}{$\alpha$=0.005} & \multicolumn{2}{|c}{$\alpha$=0.0027}\\ 
\multicolumn{2} {c|} {}&\multicolumn{2}{|c|}{(${\rm ARL}_{_0}=100$)} & \multicolumn{2}{|c|}{(${\rm ARL}_{_0}=200$)} & \multicolumn{2}{|c|}{(${\rm ARL}_{_0}=370$)} &\multicolumn{2}{|c|}{(${\rm ARL}_{_0}=100$)} & \multicolumn{2}{|c|}{(${\rm ARL}_{_0}=200$)} & \multicolumn{2}{|c}{(${\rm ARL}_{_0}=370$)}\\ 
\hline 
\multicolumn{2} {c|} {}&  {\centering{\tiny With}} & {\centering{\tiny Without}}& {\tiny With} & {\tiny Without}& {\tiny With} & {\tiny Without}& {\tiny With} & {\tiny Without}& {\tiny With} & {\tiny Without}& {\tiny With} & {\tiny Without}\\
\multicolumn{2} {c|} {} &  {\centering{\tiny CF}} & {\centering{\tiny CF}}& {\tiny CF} & {\tiny CF}& {\tiny CF} & {\tiny CF}& {\tiny CF} & {\tiny CF}& {\tiny CF} & {\tiny CF}& {\tiny CF} & {\tiny CF}\\\hline
$p$=10&Simulated ${\rm ARL}_{_1}$ &30.3&14.1 &51.1&18.5&84.6& 23.4&44.7 &15.2&79.1&19.6&129.4&23.9\\\hline 
&Nominal ${\rm ARL}_{_1}$&\multicolumn{2} {|c|} {33.2} &\multicolumn{2} {|c|} {60.1}&\multicolumn{2} {|c|} {102.3} &\multicolumn{2} {|c|} {41}&\multicolumn{2} {|c|} {75.7} &\multicolumn{2} {|c} {131.05}\\ \hline
$p$=20&Simulated ${\rm ARL}_{_1}$&20.7 &11.4 &34.7 &15.7&54.6 &20.5&27.8 &12.1 &46.3&15.5&73.8&19.5\\ \hline 
&Nominal ${\rm ARL}_{_1}$&\multicolumn{2} {|c|} {22.1} &\multicolumn{2} {|c|} {38.5}&\multicolumn{2} {|c|} {63.3} &\multicolumn{2} {|c|} {29.7}&\multicolumn{2} {|c|} {53.2} &\multicolumn{2} {|c} {89.8}\\ \hline
$p$=30&Simulated ${\rm ARL}_{_1}$&15.9&9.8  &26 &13.4&40.2 &17.8&20.9 &10.6 &33.9&13.8&52.6&17\\ \hline 
&Nominal ${\rm ARL}_{_1}$&\multicolumn{2} {|c|} {16.6} &\multicolumn{2} {|c|} {27.9}&\multicolumn{2} {|c|} {44.7} &\multicolumn{2} {|c|} {23.5}&\multicolumn{2} {|c|} {41} &\multicolumn{2} {|c} {67.8}\\ \hline
$p$=50&Simulated ${\rm ARL}_{_1}$&10.8 &7.4 &16.8 &10.1&25.5&13.4 &14.5 &8.6&23.1&11.2&33.8&13.9\\ \hline 
&Nominal ${\rm ARL}_{_1}$&\multicolumn{2} {|c|} {10.8} &\multicolumn{2} {|c|} {17.4}&\multicolumn{2} {|c|} {26.8} &\multicolumn{2} {|c|} {16.4}&\multicolumn{2} {|c|} {27.6} &\multicolumn{2} {|c} {44.3}\\ \hline
$p$=80&Simulated ${\rm ARL}_{_1}$&7.3 &5.6 &10.9 &7.4&15.6 &9.7&10.2&6.8 &15.3&8.8&22.2&11\\ \hline 
&Nominal ${\rm ARL}_{_1}$&\multicolumn{2} {|c|} {6.9} &\multicolumn{2} {|c|} {10.5}&\multicolumn{2} {|c|} {15.5} &\multicolumn{2} {|c|} {11.16}&\multicolumn{2} {|c|} {18} &\multicolumn{2} {|c} {27.8}\\ \hline
$p$=100&Simulated ${\rm ARL}_{_1}$&5.9 &4.7 &8.4 &6.2&12.2 &8&8.6 &6.1 &12.7&7.8&18.4&9.7\\ \hline 
&Nominal ${\rm ARL}_{_1}$&\multicolumn{2} {|c|} {5.5} &\multicolumn{2} {|c|} {8.1}&\multicolumn{2} {|c|} {11.7} &\multicolumn{2} {|c|} {9.1}&\multicolumn{2} {|c|} {14.3} &\multicolumn{2} {|c} {21.7}\\ \hline 
$p$=150&Simulated ${\rm ARL}_{_1}$&4 &3.35 &5.4 &4.3&7.3&5.4 &6.2&4.5 &8.8 &5.9&12.1&7.5\\ \hline 
&Nominal ${\rm ARL}_{_1}$&\multicolumn{2} {|c|} {3.6} &\multicolumn{2} {|c|} {5}&\multicolumn{2} {|c|} {6.8} &\multicolumn{2} {|c|} {6.1}&\multicolumn{2} {|c|} {9.2} &\multicolumn{2} {|c} {13.3}\\ \hline 
$p$=200&Simulated ${\rm ARL}_{_1}$&2.9 &2.6 &3.9 &3.3&5.1 &4.1&4.7 &3.8 &6.7&4.8&8.9&5.8\\ \hline 
&Nominal ${\rm ARL}_{_1}$&\multicolumn{2} {|c|} {2.7} &\multicolumn{2} {|c|} {3.5}&\multicolumn{2} {|c|} {4.6} &\multicolumn{2} {|c|} {4.6}&\multicolumn{2} {|c|} {6.5} &\multicolumn{2} {|c} {9.2}\\ \hline
\end{tabular}
\end{table}

\begin{figure}[h!]
  \centering
  \begin{subfigure}[b]{0.45\linewidth}
    \includegraphics[width=\linewidth, height=2.5in]{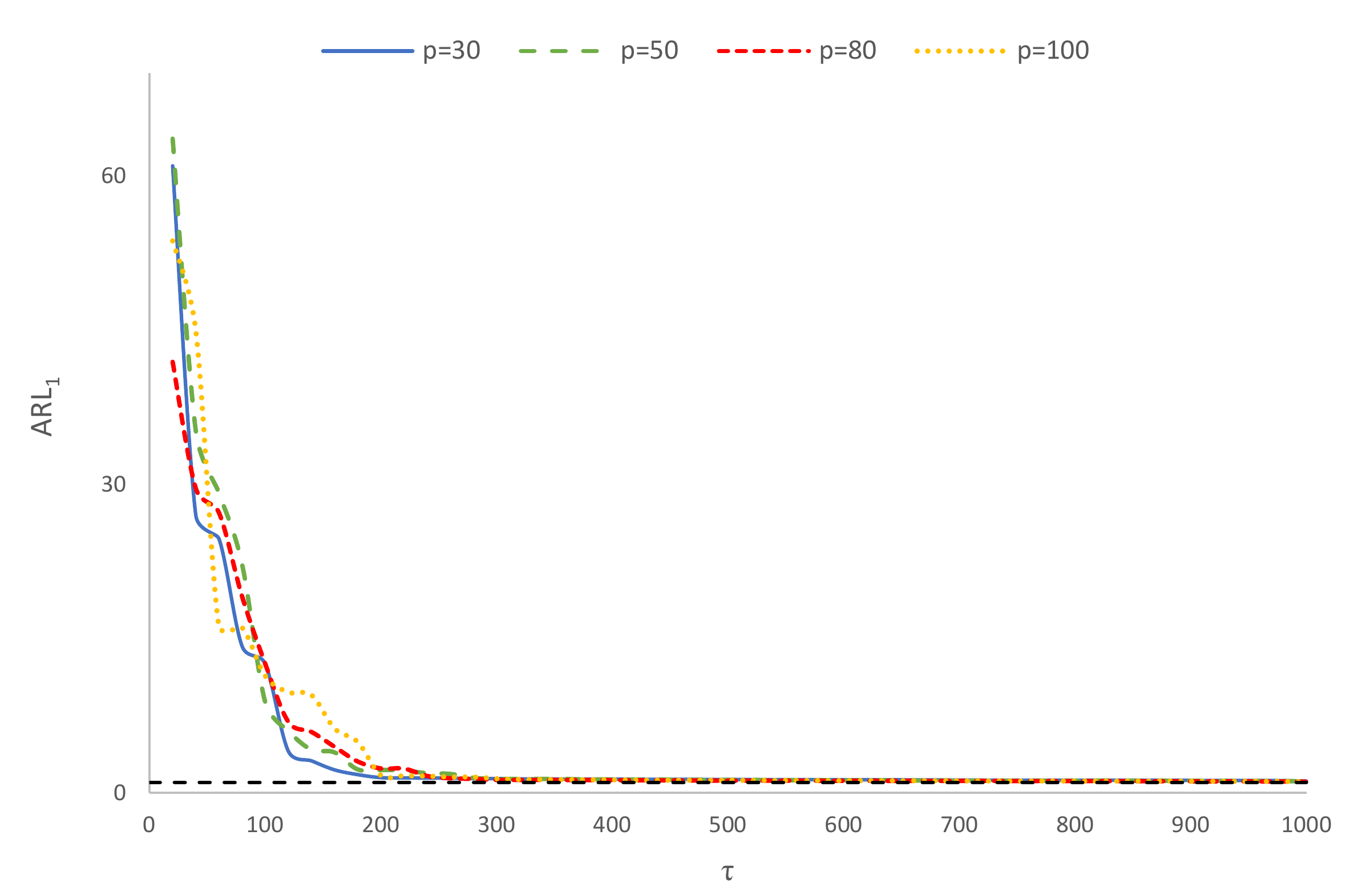}
    \caption{Scenario 1 }
  \end{subfigure}
    \begin{subfigure}[b]{0.45\linewidth}
    \includegraphics[width=\linewidth, height=2.5in]{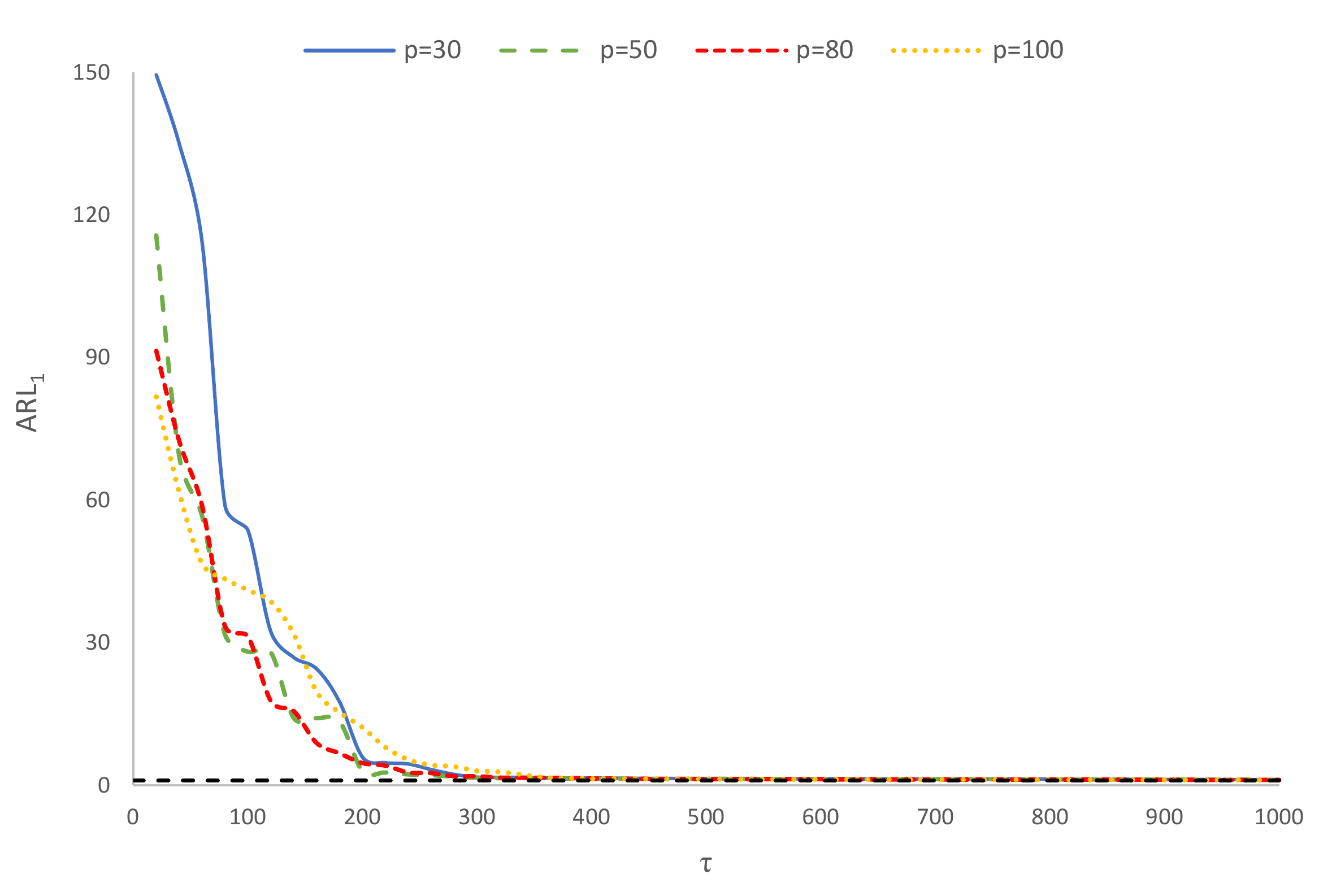}
    \caption{Scenario 2}
  \end{subfigure}
  \caption{A comparison of the simulated ${\rm ARL}_{_1}$ (coloured curves) in a self starting chart with nominal ${\rm ARL}_{_1}$ (horizontal dash lines) for different values $p$ and $\tau$.}
  \label{fig:self}
\end{figure}

\begin{table}[h!]

\tiny
\centering
\caption{Effect of correlation on ${\rm ARL}_{_0}$ of the proposed chart (cavariance matrix $\sigma_{ij}=(a)^{|i-j|}\quad {\rm for}\quad i,j=1,\dots,p$ with different values of $a$)}
\label{table:cor}
\begin{tabular}{ |p{1.1cm}|p{0.85cm}|p{0.85cm}|p{0.85cm}|p{0.85cm}|p{0.85cm}|p{0.85cm}|p{0.85cm}|p{0.85cm}|p{0.85cm}|p{0.85cm}|}
 \hline
\multicolumn{11} {|c|} {$\alpha$=0.01 (${\rm ARL}_{_0}$=100)}\\ \hline
{\centering{$a$}}& {\centering{0}} & {\centering{0.1}} & {\centering{0.2}}& {0.3} & {0.4}& {0.5} & {0.6}& {0.7} & {0.8}& {0.9}\\\hline
$p$=30&103&101 &102&100&102&102&104 &107&112&124\\ \hline 
$p$=50&102 &103 &97 &101&98 &101&100 &102 &105&115\\ \hline 
$p$=80&101&101 &99 &100&102 &103&98 &103 &106&111\\ \hline 
$p$=100&101&94&102  &101 &99&101 &102&101 &101 &106\\ \hline
\multicolumn{11} {|c|} {$\alpha$=0.005 (${\rm ARL}_{_0}$=200)}\\ \hline
{\centering{$a$}}& {\centering{0}} & {\centering{0.1}} & {\centering{0.2}}& {0.3} & {0.4}& {0.5} & {0.6}& {0.7} & {0.8}& {0.9}\\\hline
$p$=30&202&202 &200&192&200&202&208 &196&209&224\\ \hline 
$p$=50&204 &209 &194 &197&192 &198&190 &203 &203&210\\ \hline 
$p$=80&201&199 &203&203 &193&195 &196&194 &197 &213\\ \hline 
$p$=100&202&205  &199 &191&202 &197&200 &197 &196&206\\ \hline
\end{tabular}
\end{table}
It is common in the literature to judge the performance of a self-starting chart by its out-of-control run lengths. See for example, Hawkins and Maboudou-Tchao (2007), Zou et al. (2007) and Maboudou-Tchao and Hawkins (2011). Several factors such as dimension, in-control ARL, size of the shift in the mean vector may affect the performance of a self-starting chart. An important factor, especially in the unknown-parameter self-starting setting, is the ``learning time” of a chart, which is the length of time the process runs in control before a shift occurs (Maboudou-Tchao and Hawkins 2011). A chart with shorter learning times is highly desired. We now evaluate the effect of an initial in-control period of length $\tau$ on the out-of-control ARL performance of the proposed self-starting chart for different dimensions of the process. Figure \ref{fig:self} displays the effect of the initial in-control period of length $\tau$ on ${\rm ARL}_{_1}$ for $p=30,\,50,\,80,\,100$. We note that asymptotic type II error probability of proposed chart is through the parameter $\eta$ as
\begin {equation}
\eta=\frac{{\bfmath{\delta}}^{\prime}\mathbf{D}^{-1}\bfmath{\delta}}{\sqrt{2\,{\rm tr}(\bfmath{\rho}^{2})}}\,,
\end {equation}
where $\bfmath{\mu}_{_1}-\bfmath{\mu}=\bfmath{\delta}$.

In figure \ref{fig:self}, it is assumed that new (out-of-control) observations have the mean vector $\bfmath{\mu}_1$ which is determined such that $\eta=5$ in all cases. The ${\rm ARL}_{_0}$ is set to 200, and the nominal ${\rm ARL}_{_1}$ for these choices of $\eta$ and $\alpha$ is 1.0077, calculated based on equation \eqref{arls}, for all values of $p$ in both Scenarios 1 and 2. This asymptotic value is also shown as a horizontal dash line in Figure \ref{fig:self}. A selection of $\tau$ values ranging from 20 to 1000 is used. After an initial in-control period of length $\tau$, the mean vector was shifted, and all signals before time $\tau$ were omitted from the calculations. The figure shows that generally ${\rm ARL}_{_1}$ decreases slightly with increases in $\tau$. In other words, when the initial learning period is short, the chart may take a long time to detect the shift, but by increasing $\tau$, the detection time improves significantly. The figure also shows that the proposed chart reaches it's asymptotic ${\rm ARL}_{_1}$ between $\tau=200$ and 300 for all dimensions in both scenarios.

\begin{figure}[h!]
  \centering
  \begin{subfigure}[b]{0.32\linewidth}
    \includegraphics[width=\linewidth, height=2.3in]{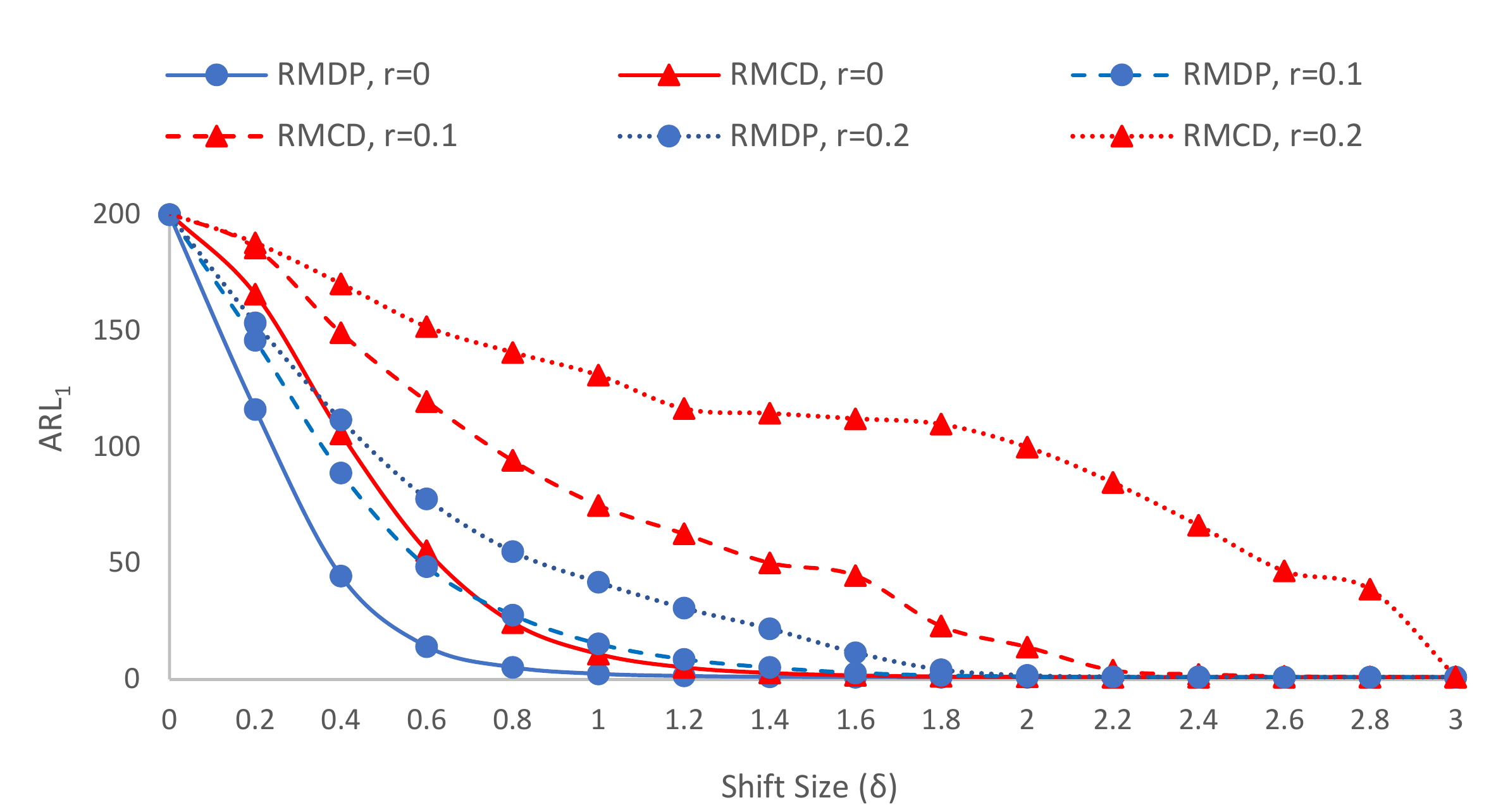}
    \caption{$p=30$ }
  \end{subfigure}
  \begin{subfigure}[b]{0.32\linewidth}
    \includegraphics[width=\linewidth, height=2.3in]{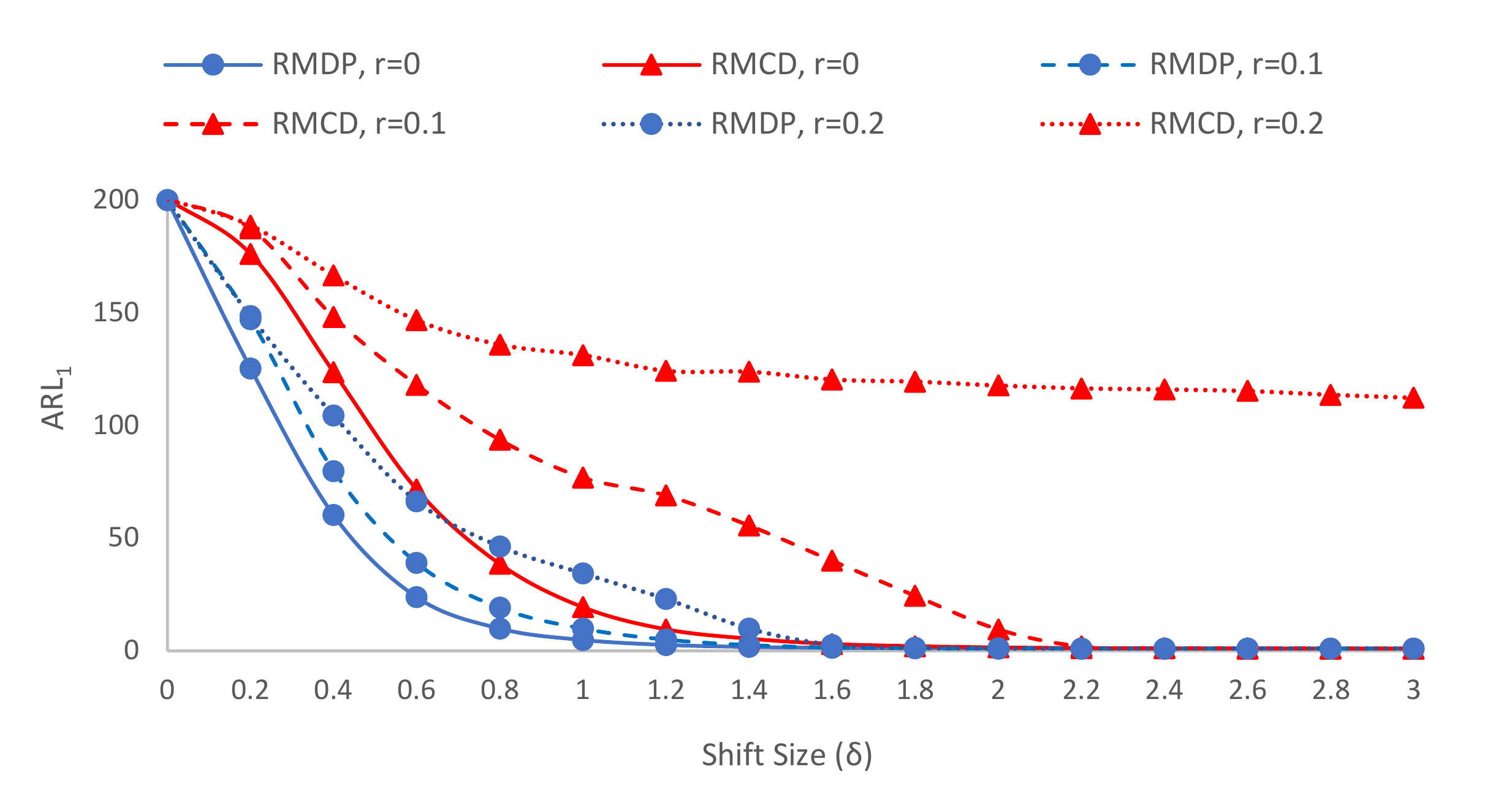}
    \caption{$p=50$ }
  \end{subfigure}
  \begin{subfigure}[b]{0.32\linewidth}
    \includegraphics[width=\linewidth, height=2.3in]{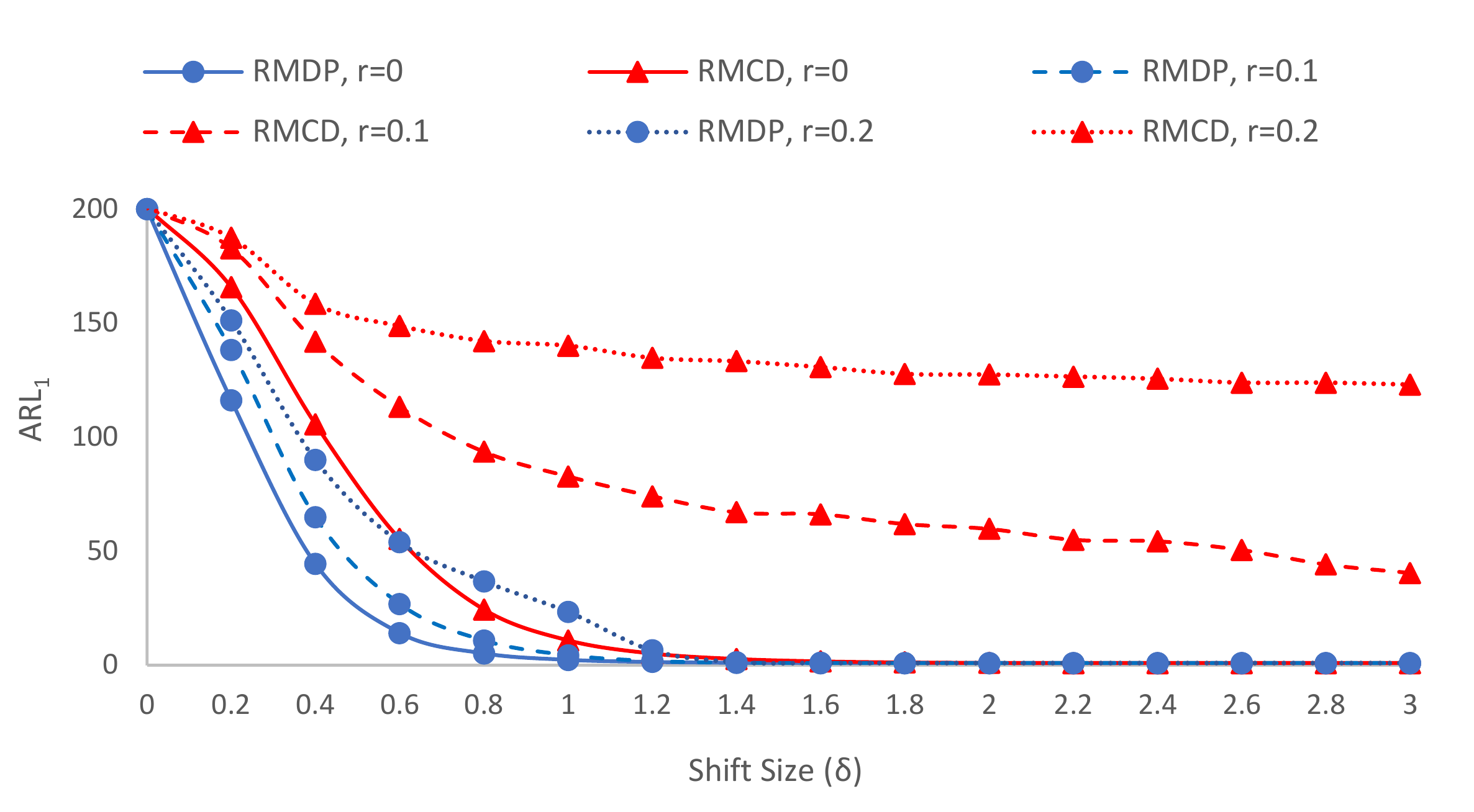}
    \caption{$p=100$}
  \end{subfigure}
  \caption{A comparison between ${\rm ARL}_{_1}$s of the proposed RMDP chart and the RMCD chart of Chenouri et al. (2009) in Phase II when 50\% of variables are shifted in an amount of $\delta$ in both Phase I and Phase II and the desired false-alarm rate is 0.005.}
  \label{fig7}
\end{figure}

To provide a better understanding of the proposed control chart's performance in Phase II, we compare it with the RMCD control chart proposed by Chenouri et al. (2009). Based on using RMCD estimators as robust estimators of the mean vector and covariance matrix, Chenouri et al. (2009) proposed a robust Hotelling's $T^2$-type control chart for individual observations. Comparing our method with the aforementioned RMCD chart is sensible because both methods are Shewhart-type control charts and use robust approaches and reweighting algorithms. Chenouri et al. (2009) used an extensive simulation to estimate the empirical $99\%$ and $99.9\%$ quantiles of Phase II $T^2$ chart when the Phase I sample size $m$ is not large. However, they did not provide the empirical quantiles for large dimension $p$. It is worth mentioning that a significant advantage of our proposed chart is the use of Cornish-Fisher expansion and the finite sample correction coefficient described in Section 2.1. We do not estimate the quantiles of the charting statistic for different values of $p$, which can be a very time-consuming task especially when $p$ is large. So, to perform a fair comparison, we assume that $m$ is 100,000. We design our experiment so that both methods theoretically achieve ${\rm ARL}_{_0}=200$ when $\alpha=0.005$. Our proposed RMDP method and the RMCD method of Chenouri et al. (2009) use their respective parameters' estimates in Phase I. The function {\tt CovMcd} in the {\tt rrcov} package of {\tt R} software written by Valentine Todorov (2007) is used to calculate process parameters based on RMCD. We compare the performance of our proposed control chart (without the adaptive feature) with the RMCD chart in Phase II in terms of ${\rm ARL}_{_1}$ for different amounts of shift. We consider three different rates of contamination $r=0, \,0.1,\,0.2$ in Phase I data. For the sake of simplicity, we assume that the mean vector of the contaminating distribution in Phase I is the same as that of Phase II out-of-control observations. Note that the non-centrality parameter of the $T^2$ chart is different from $\eta$. So, for any Phase I outlier or Phase II out-of-control observation, we assume that the mean of the first $p_{_1}=0.3\,p, \,0.5\,p, 0.8\,p$ variables are equally shifted by the amount of $\delta=0.2, \,0.4, \,\,\dots ,\,3$ while the covariance matrix remains in-control. In Phase I, out of the generated $m$ observations, $\lfloor m\,r\rfloor$ of them are outliers with distribution ${\rm N}(\bfmath{\mu}_{_1},\,\bfmath{\Sigma})$ and the remaining $\lfloor m\,(1-r)\rfloor$ observations are generated from the in-control distribution ${\rm N}(\bfmath{\mu},\,\bfmath{\Sigma})$. In Phase II, the new out-of-control observations with distribution ${\rm N}(\bfmath{\mu}_{_1},\,\bfmath{\Sigma})$ are generated to compute ${\rm ARL_{_1}}$ for both methods based on 10,000 simulations. We only report the comparison between the two methods for Scenario 2, while we can make similar conclusions for Scenario 1. Figure \ref{fig7} depicts the comparison of our proposed RMDP chart with the RMCD chart of Chenouri et al. (2009) for some combinations of $p$, $m$, and $r$ when $p_{_1}=0.5\,p$ and $\alpha=0.005$. The curves related to RMDP and RMCD are in blue and red colours, respectively. Looking at  Figure \ref{fig7}, we conclude that our proposed method outperforms the RMCD chart for fixed $r$ and $p$ in terms of ${\rm ARL}_{_1}$. When $r=0$, the ${\rm ARL}_{_1}$ from both methods converges to 1 for any $p$ as the mean shift value increases. The proposed RMDP chart's performance in Phase II does not change considerably when contaminations exist among Phase I data. However, the ${\rm ARL}_{_1}$ of the RMCD based method stays far from 1 for large $\delta$ when $r=0.1$ and $0.2$, especially when $p$ increases. Moreover, when $r=0$, the simulated ${\rm ARL}_{_1}$s for RMCD method for different shifts are very close to their theoretical values calculated based on the non-central Chi-square distribution. For $r=0.1$ and $0.2$, the simulated ${\rm ARL}_{_1}$s of RMCD chart are bigger than their theoretical values. The simulation results for $p_{_1}=0.3\,p$ and $p_{_1}=0.8\,p$ support the conclusions mentioned above. 

\section{A real world example data}
 
In this Section, we provide an example of Phase II monitoring using a multivariate dataset for a semiconductor manufacturing
process. The dataset was recently used by Zou et al. (2015), Shu and Fan (2018), Li et al. (2020), and Mukherjee and Marozzi (2020) for high-dimensional monitoring and is available online at the UC Irvine Machine Learning Repository
\url{(http://archive.ics.uci.edu/ml/datasets/SECOM) }.  This dataset was
collected from July 2008 to October 2008,  consists of  1567 vector observations, and for each observation, there are 591 continuous measurements (variables). However, the dataset contains a considerable number of null (missing) values and several variables with almost constant values. After cleaning data, in total $p=250$ variables remained. Among 1567 vector observations in the dataset, 1463 observations are labelled as conforming, and the remaining 104 as non-conforming. There is also a label that specifies the timestamp of each sample. The abovementioned papers, which investigated this dataset,  treated the vector of observations related to conforming parts as in-control Phase I data. They then monitored non-conforming items as Phase II data based on the estimated parameters from the Phase I data. However, a more reasonable approach is to divide the data into Phase I and Phase II data based on their sampling time. We consider the first 80\% of data as historical (Phase I) data, while the next 20\% as Phase II data for monitoring. In other words, $\mathbf{X}_1,\dots,\mathbf{X}_{1253}$ are Phase I data and $\mathbf{X}_{1254},\dots,\mathbf{X}_{1567}$ are for Phase II monitoring. Our objective is estimating process parameters based on Phase I data and then monitoring the Phase II observations based on the proposed control chart and estimated parameters from Phase I. 
 
Ignoring the correlation among variables, we conduct marginal normality tests of Shapiro–Wilks. We conclude that the assumption of normality does not hold for most of the variables (p-values are very small). So, for each marginal observation $X_{ij},\quad i=1,...,250,\quad j=1, ...,1567$, we use the inverse transformation $\Phi^{-1}(\widehat{F}_{i}(X_{ij})),$ where $\widehat{F}_{i}$ is the marginal empirical distribution function based on the 1463 conforming observations of the $i$th variable. 

\begin{figure}[h!]
  \centering
    \begin{subfigure}[b]{1\linewidth}
    \includegraphics[width=\linewidth]{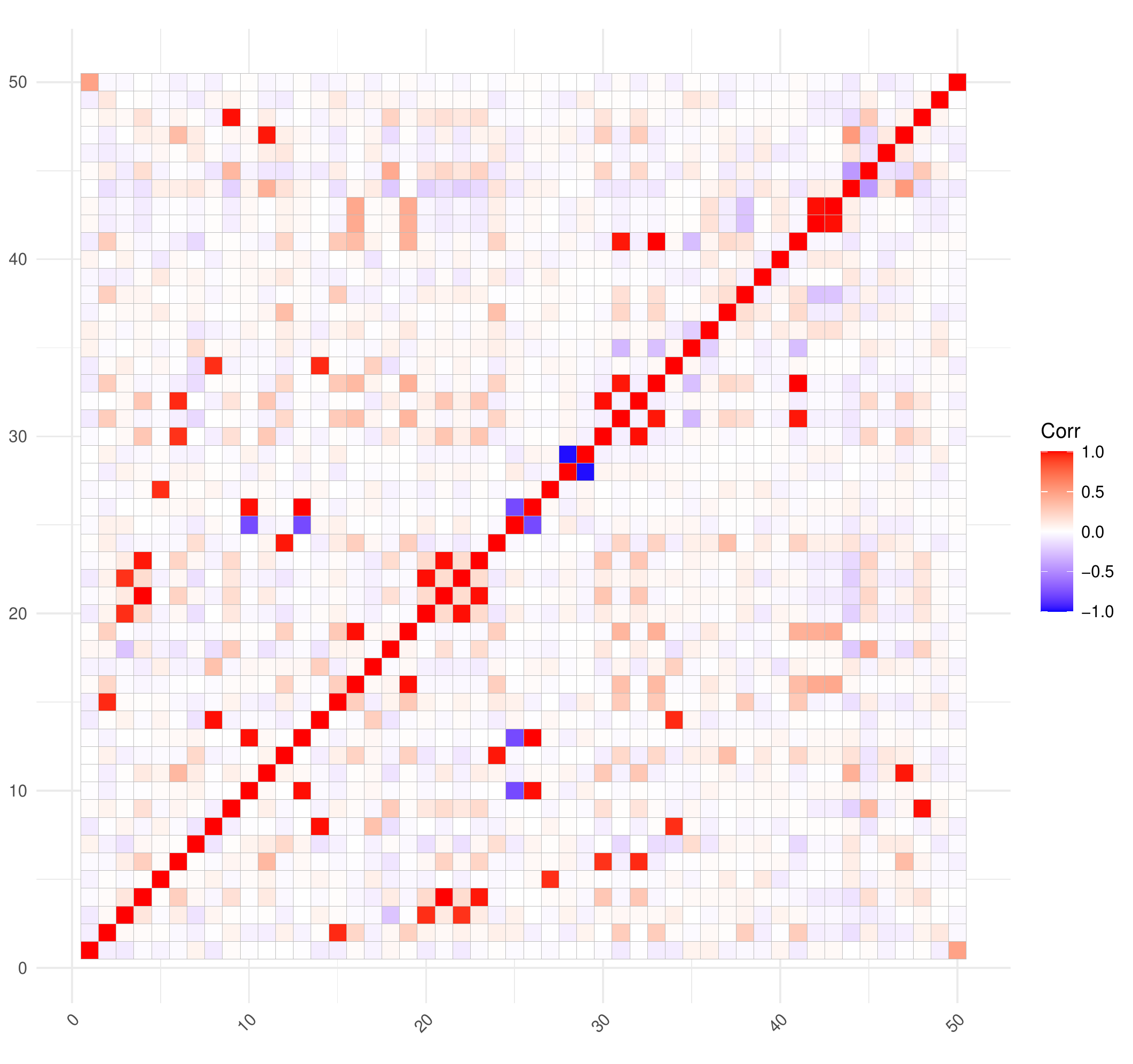}
  \end{subfigure}  
  \caption{Correlation between first 50 variables of semiconductor manufacturing data.}
  \label{fig10}
\end{figure}

\begin{figure}[h!]
  \centering
    \begin{subfigure}[b]{1\linewidth}
    \includegraphics[width=\linewidth]{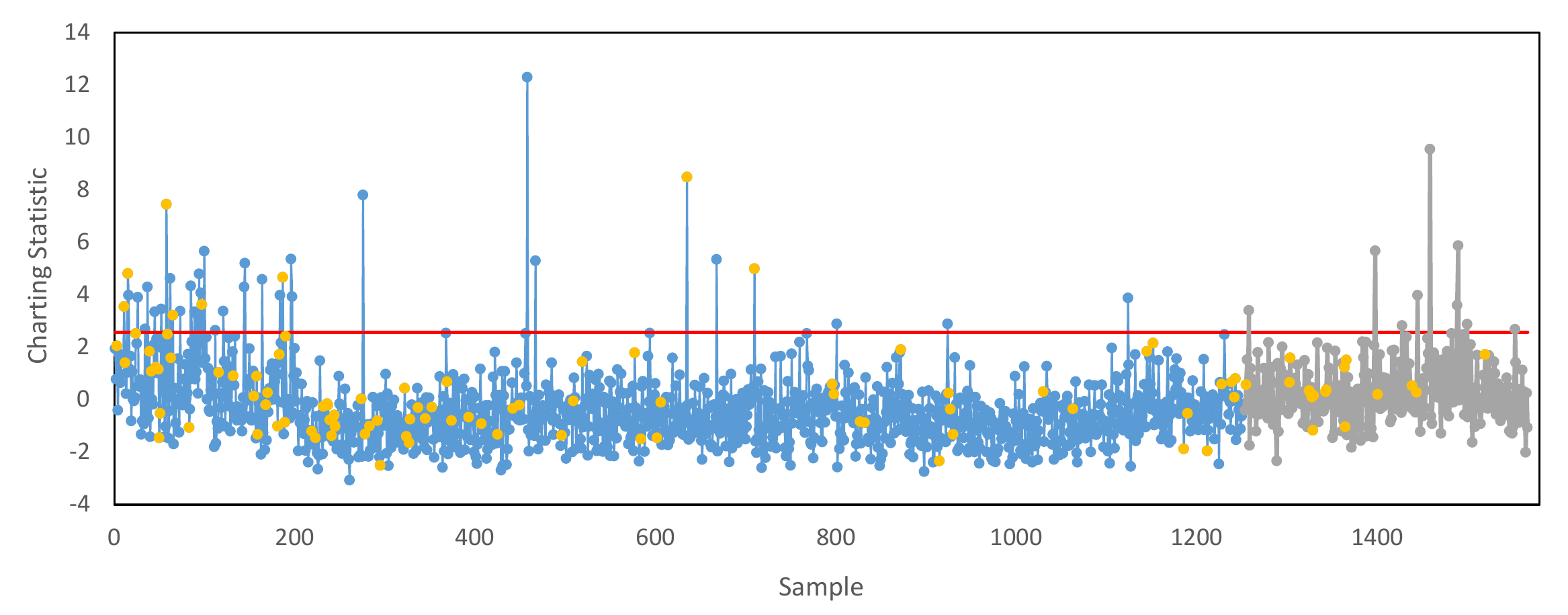}
  \end{subfigure}  
  \caption{A control chart when $80\%$ of samples are considered in Phase I (blue points) and $20\%$ in Phase II (grey points) for semiconductor data.}
  \label{fig11}
\end{figure}

\begin{figure}[h!]
  \centering
    \begin{subfigure}[b]{1\linewidth}
    \includegraphics[width=\linewidth]{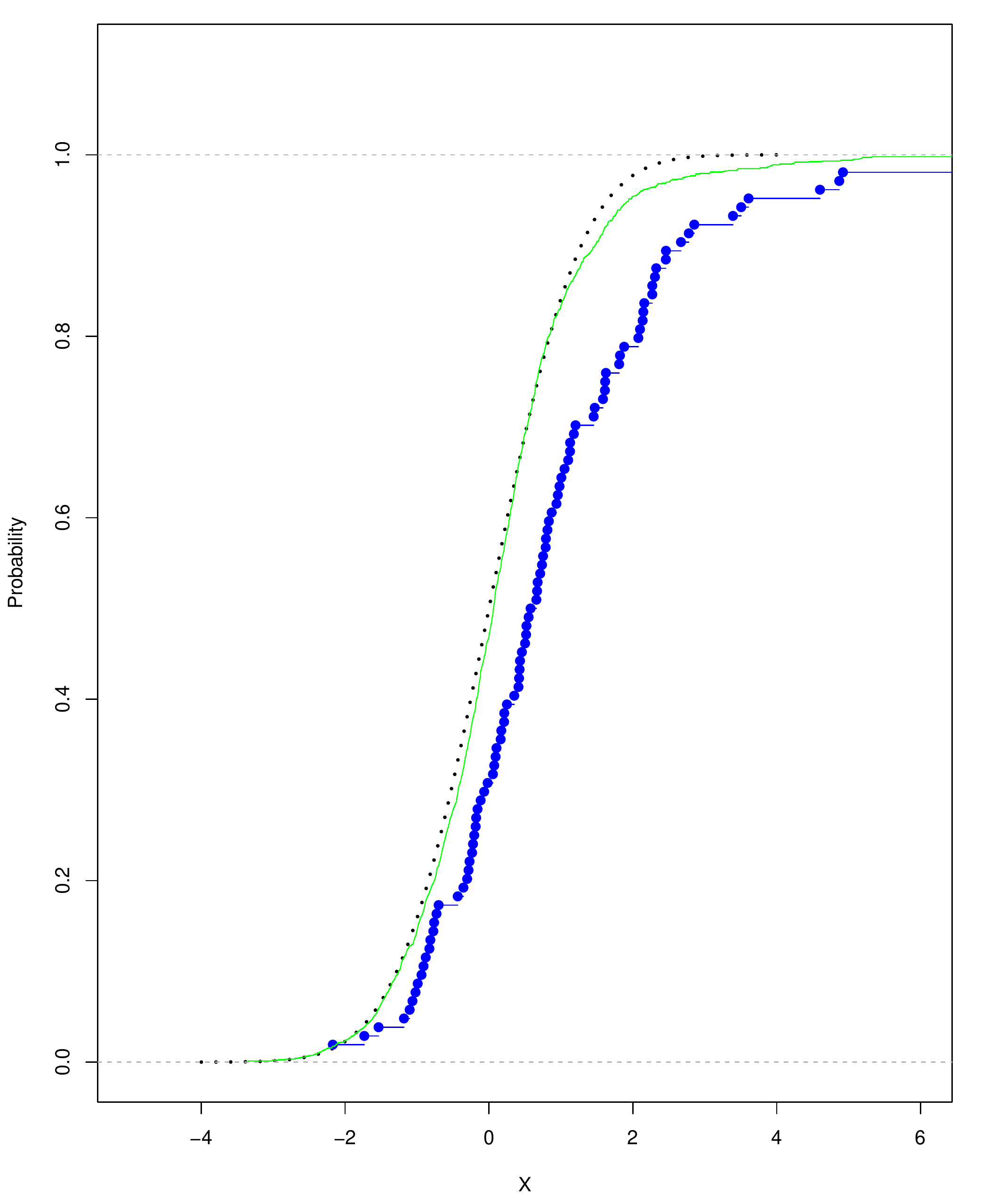}
  \end{subfigure}  
  \caption{A comparison between empirical CDF of charting statistics of conforming samples (green) and non-conforming samples (blue) with the CDF standard normal (dashed black line).}
  \label{fig12}
\end{figure}

We implement the proposed robust RMDP approach on 1253 samples of Phase I. We consider $\alpha=0.005$ to achieve $ARL_0$=200, so the estimated parameters from Phase I are  $c^{^{\rm MDP}}_{p,m}=1.01$, $\widehat{{\rm tr}(\bfmath{\rho}^2)}_{_{\rm MDP}}=908$, and $\widehat{{\rm tr}(\bfmath{\rho}^3)}_{_{\rm MDP}}=5817$. Figure \ref{fig10} shows that the estimated correlation matrix based on Phase I samples is sparse, while some variables are highly correlated. Hence, though the pairwise correlations among most variables are weak, some strong correlations are consistent with our
assumptions based on our previous discussion in Section 2.1. After estimating the required parameters, we construct the propsed control chart to examine which of 314 samples in Phase II are out of control. Figure \ref{fig11} depicts the control chart for Phase II observations based on our proposed method, where the UCL is $z_{0.005}=2.575$. In Figure \ref{fig11}, Phase I data, Phase II data, and non-conforming items are specified by blue, grey, and yellow points, respectively. Figure \ref{fig11} suggests a shift in the process mean of the first 200 observations in Phase I, which implies that the proposed chart can quickly detect and declare the process as out-of-control. It is worthwhile to mention that among 104 non-conforming items, 29 appear within the first 200 items.  This observation suggests that an initial investigation of the process could have prevented the subsequent out-of-control or non-conforming items. 
Looking at Figure \ref{fig11}, we see an apparent second shift in the process mean appearing in Phase II data from sample 1253 to 1567, in which there are 17  non-conforming items.  However, similar to the distribution-free control charts proposed by Shu and Fan (2018) and Mukherjee and Marozzi (2020), for most of the nonconforming items, the value of the charting statistics is below the calculated UCL. This contradictory result might be because non-conforming items are not due to the process mean shifts only. Figure \ref{fig12} compares the empirical CDF of the charting statistics of conforming and non-conforming samples with that of the standard normal. While the empirical CDF of conforming samples shows a perfect match with the standard normal CDF, the empirical CDF of non-conforming items represents a shift to the right from the standard normal distribution, revealing a noticeable change in distribution.

\section{Conclusion}
In many practical applications of high dimensional processes, the Phase I sample size is usually small and computing the sample covariance matrix is impractical.  In this article, we employ the diagonal matrix of the underlying covariance matrix to monitor the mean vector of high dimensional correlated quality characteristics described by a multivariate normal distribution. The main reason to consider the diagonal matrix instead of the whole sample covariance matrix is its non-singularity for cases where the number of quality characteristics is much larger than the sample size. Moreover, we proposed a unifying approach for Phase I and Phase II analysis by employing a self-starting control chart. In terms of efficiency, the proposed procedure shows good performance in Phase II.  
Due to recent advances in data-acquisition equipment, the study of high dimensional process monitoring is a vibrant and promising research area. We believe that much more work is needed on this topic.

\section*{References}
\begin{enumerate}
\item[] Abramowitz, M., Stegun, I. A. (Eds.). 1972. {\it Handbook of mathematical functions with formulas, graphs, and mathematical tables} (Vol. 55). Washington, DC: National bureau of standards.
\item[] Abdella, G.M., Al‐Khalifa, K.N., Kim, S., Jeong, M.K., Elsayed, E.A. and Hamouda, A.M., 2016. Variable Selection‐based Multivariate Cumulative Sum Control Chart. {\it Quality and Reliability Engineering International, 33}, 565-578.
\item[] Ahmadi-Javid, A., Ebadi, M. (2021). A two-step method for monitoring normally distributed multi-stream processes in high dimensions. {\it Quality Engineering, 33}, 143-155.
\item[] Bersimis, S., S. Psarakis, and J. Panaretos. 2007. Multivariate Statistical Process Control Charts: An Overview. {\it Quality and Reliability Engineering International, 23}, 517–543.
\item[] Capizzi, G., Masarotto, G. 2011. A least angle regression control chart for multidimensional data. {\it Technometrics, 53},  285–296.
\item[] Chen, N., Zi, X., Zou, C. 2016. A distribution-free multivariate control chart. {\it Technometrics, 58}, 448-459.
\item[] Chenouri, S. E., Steiner, S. H.,  Variyath, A. M. 2009. A multivariate robust control chart for individual observations. {\it Journal of Quality Technology, 41}, 259-271.
\item[] Cornish, E. A., Fisher, R. A. 1938. Moments and cumulants in the specification of distributions.  {\it Review of the International Statistical Institute, 5}, 307-320.
\item[] DasGupta, A. 2008. {\it Asymptotic theory of statistics and probability}. Springer Science \& Business Media.
\item[] Ebadi, M., Chenouri, S., Steiner, S. H. 2021. Phase I Analysis of High-Dimensional  Processes in the Presence of Outliers. To appear in {\it Journal of Quality Technology}. {\it Arxiv Preprint arXiv:2110.13689}.
\item[] Fisher, S. R. A., Cornish, E. A. 1960. The percentile points of distributions having known cumulants. {\it Technometrics, 2}, 209-225.
\item[] Hall, P. 1983. Inverting an Edgeworth expansion. {\it The Annals of Statistics, 11}, 569-576.
\item[] Hawkins, D. M., and Maboudou-Tchao, E. M. 2007. Self-starting multivariate exponentially weighted moving average control charting. {\it Technometrics, 49}, 199-209.
\item[] Jiang, W., Wang, K., Tsung, F. 2012. A variable-selection-based multivariate EWMA chart for process monitoring and diagnosis. {\it Journal of Quality Technology, 44}, 209–230.
\item[] Kim, S., Jeong, M. K., \& Elsayed, E. A. 2020. A penalized likelihood-based quality monitoring 
via L2-norm regularization for high-dimensional processes. {\it Journal of Quality Technology, 52}, 
265-280.
\item[] Li, W., Xiang, D., Tsung, F., Pu, X. 2020. A diagnostic procedure for high-dimensional data streams via missed discovery rate control. {\it Technometrics, 62}, 84-100.
\item[] Maboudou-Tchao, E. M., Hawkins, D. M. 2011. Self-starting multivariate control charts for location and scale. {\it Journal of Quality Technology, 43}, 113-126.
\item[]
Mahalanobis P.C. 1936,
On the generalised distance in statistics,
{\it Proceedings of the National Institute of Science of India, 12}, pp. 49-55
\item[] Mukherjee, A., \& Marozzi, M. 2020. Nonparametric Phase-II control charts for monitoring high-dimensional processes with unknown parameters. {\it Journal of Quality Technology}, 1-21.
\item[] Polansky, A. M. 2011. {\it Introduction to statistical limit theory}. CRC Press.

\item[] Quesenberry, C. P. 1997. {\it SPC methods for quality improvement.} New York: Wiley.
\item[] Ro, K., Zou, C., Wang, Z., \& Yin, G. 2015. Outlier detection for high-dimensional data. {\it Biometrika, 102}, 589-599.
\item[] Satterthwaite, F. E. 1941. Synthesis of variance. {\it Psychometrika, 6}, 309-316.
\item[] Satterthwaite, F. E. 1946. An approximate distribution of estimates of variance components. {\it Biometrics bulletin, 2}, 110-114.
\item[] Shu, L., Fan, J. 2018. A distribution‐free control chart for monitoring high‐dimensional processes based on interpoint distances. {\it Naval Research Logistics (NRL), 65}, 317-330.
\item[] Small, C. G. 2010. {\it Expansions and asymptotics for statistics.} CRC Press.
\item[] Srivastava, M. S., 2005. Some tests concerning the covariance matrix in high dimensional data. {\it Journal of the Japan Statistical Society, 35}, 251-272.
\item[] Srivastava, M. S., and Du, M. 2008. A test for the mean vector with fewer observations than the dimension. {\it Journal of Multivariate Analysis, 99}, 386-402.
\item[] Srivastava, M. S., Katayama, S., Kano, Y. 2013. A two sample test in high dimensional data. {\it Journal of Multivariate Analysis, 114}, 349-358.
\item[] Srivastava, M. S., Yanagihara, H. 2010. Testing the equality of several covariance matrices with fewer observations than the dimension. {\it Journal of Multivariate Analysis, 101}, 1319-1329.
\item[] Sullivan, J. H., Jones-Farmer, L. A. 2002. A self-starting control chart for multivariate individual observations. {\it Technometrics, 44}, 24-33.

\item[] Wang, K., Jiang, W. 2009. High-dimensional process monitoring and fault isolation via
variable selection. {\it Journal of Quality Technology, 41}, 247–258.
\item[] Welch, B. L. 1951. On the comparison of several mean values: an alternative approach. {\it Biometrika, 38}, 330-336.
\item[] Woodall, W. H., Montgomery, D. C. 2014. Some current directions in the theory and application of statistical process monitoring. {\it Journal of Quality Technology, 46}, 78-94.
\item[] Zhang, L., Zhu, T., Zhang, J. T. 2020. A simple scale-invariant two-sample test for high-dimensional data. {\it Econometrics and Statistics, 14}, 131-144.
\item[] Zou, C., Qiu, P. 2009. Multivariate statistical process control using LASSO. {\it Journal of
American Statistical Association, 104}, 1586–1596.
\item[] Zou, C., Wang, Z., Zi, X., Jiang, W. (2015). An efficient online monitoring method for high-dimensional data streams. {\it Technometrics, 57}, 374-387.
\item[] Zou, C., Zhou, C., Wang, Z., Tsung, F. 2007. A self-starting control chart for linear profiles. {\it Journal of Quality Technology, 39}, 364-375.
\end{enumerate}

\appendix
\newpage
\section*{Appendix A}
In this Appendix, we derive the Cornish-Fisher expansion in \eqref{CFexp} for the proposed test statistic $U_{_i}$. We first briefly review the Cornish-Fisher expansion approach. For more details, we refer the reader to Fisher and Cornish (1960), Abramowitz and Stegun (1972), and Polansky (2011). 
Let $Y_1,\,\cdots ,\,Y_{n}$ be a sequence of independent and identically distributed random variables with mean $m$ and variance $\sigma^2$. Denote the the cumulative distribution function of $Y=\sum\limits_{i=1}^{n}Y_i$ by $F_{n}(y)$. Let $y_{_\alpha}$ represent the $(1-\alpha)100\%$ quantile of $F_{n}$, that is $F_{n}(y_{_{\alpha,\,n}})=1-\alpha$. The Cornish-Fisher asymptotic expansion of $y_{_{\alpha,\,n}}$ with respect to $n$ is given by $y_{_{\alpha,\,n}}\sim m+\sigma\,\omega_{_{\alpha,\,n}}$, where
\begin{equation}\label{C-F}
\omega_{_{\alpha,\,n}}=z_{_{\alpha}}+\left[\,\gamma_{_1}\,h_{_1}(z_{_{\alpha}})\,\right]+\left[\,\gamma_{_2}\,h_{_2}(z_{_{\alpha}})+\gamma_{_1}^2\,h_{_{11}}(z_{_{\alpha}})\,\right]+\cdots,
\end{equation}
with $z_{_{\alpha}}=\Phi^{-1}(1-\alpha)$, and 
$\gamma_{_{r-2}}=\kappa_{_2}^{-\frac{r}{2}}\,\kappa_{_r}$, for $r=3,4,\dots$, is the standardized version of the cumulants 
$\kappa_r$ of $F_{n}$. For instance, $\gamma_{_1}$ is the skewness, $\gamma_{_2}$ is the (excess) kurtosis, etc. Also,
\begin{equation}\label{e7}
h_{_1}(x)=\frac{1}{6}\,H_{_2}(x), \quad h_{_2}(x)=\frac{1}{24}\,H_{_3}(x), \quad h_{_{11}}(x)=\frac{-1}{36}\,\left[\,2\,H_{_3}(x)+H_{_1}(x)\,\right],\quad \dots\,,
\end{equation}
where $H_{_k}(x)$ is a $k^{\rm th}$-order polynomial in $x$ called the $k^{\rm}$ Hermite polynomial defined by
\begin{equation}\label{e7}
H_{_k}(x)=\sum\limits_{i=0}^{\left\lfloor k/2\right\rfloor}(-1)^i\,\frac{(2\,i)!}{2^{i}\,i!}\,{k\choose 2\,i} \,x^{k-2\,i}\,\,,
\end{equation}
where $\lfloor a\rfloor$ is greatest integer less than or equal to $a$. For example $H_{_1}(x)=x,\, H_{_2}(x)=x^2-1$ and $H_{_3}(x)=x^3-3\,x$. Abramowitz and Stegun (1972) provided an auxiliary table, in which the values of $h_{_1}(z_{_{\alpha}}), \,h_{_2}(z_{_{\alpha}}),\, \dots$,  are tabulated for some $\alpha$.

To establish the Cornish-Fisher expansion for our statistic $U_{i}$, recall that for independent and identically distributed standard normal random variables $\xi_{_1}, \,\xi_{_2},\, \dots,\, \xi_{_p}$, the test statistic can be written as
\begin{align*}
U_{i}=\frac{M^{2}_{i}(\bfmath{\mu},\mathbf{D_{_0}})-p}{\sqrt{2\,\textrm{tr}(\bfmath{\rho}^2)}}=\frac{\sum\limits_{j=1}^{p}\lambda_{_j}\xi_{_j}^{2}-p}{\sqrt{2\,\textrm{tr}(\bfmath{\rho}^2)}}
\end{align*}
Recall also that the moments of the random variables $\xi_{_j}$ are 
\begin{equation}\label{Nmoments}
{\rm E}[\,\xi_{j}^k\,]=0 \quad \text{ for } k \text{ odd}\,,\qquad  \quad {\rm E}[\,\xi_{j}^k\,]=\frac{k!\,2^{-k/2}}{\left(k/2\right)!} \quad\text{ for } k \text{ even}\,.
\end{equation}
Since $\kappa_{_1}={\rm E}\left[U\right]=0$ and $\kappa_{_2}={\rm Var}\left[U\right]=1$, to obtain the 2nd order Cornish-Fisher expansion using \eqref{C-F}, we must calculate the third and fourth cumulants, $\kappa_{_3}$ and $\kappa_{_4}$, of the statistic $U_i$. Since 
\begin{align}\label{sums}
{\rm tr}(\bfmath{\rho}^k)&=\sum\limits_{j=1}^{p}\lambda_{_j}^k \quad \text{ for any } \quad k=1,\, 2,\, \dots\,,\nonumber\\
\sum\limits_{i\neq{j}}\lambda_{i}\,\lambda_{j}&=\sum\limits_{j=1}^{p}\lambda_{j}\,\sum\limits_{j=1}^{p}\lambda_{j}-\sum\limits_{j=1}^{p}\lambda_{j}^2=p^2-\textrm{tr}(\bfmath{\rho}^2)\,, \nonumber\\
\sum\limits_{i\neq{j}}\lambda_{i}^2\,\lambda_{j}&=\sum\limits_{j=1}^{p}\lambda_{j}^2\,\sum\limits_{j=1}^{p}\lambda_{j}-\sum\limits_{j=1}^{p}\lambda_{j}^3=p\,\textrm{tr}(\bfmath{\rho}^2)-\textrm{tr}(\bfmath{\rho}^3)\,,\nonumber\\
\sum\limits_{i\neq{j}\neq{k}}\lambda_{i}\,\lambda_{j}\lambda_{k}&=\left(\sum\limits_{j=1}^{p}\lambda_{j}\right)^3-\sum\limits_{j=1}^{p}\lambda_{j}^3-3\sum_{i\neq{j}}\lambda_{i}^2\,\lambda_{j}=p^3+2\,\textrm{tr}(\bfmath{\rho}^3)-3\,p\,\textrm{tr}(\bfmath{\rho}^2)\,,\nonumber\\  
\sum\limits_{i\neq{j}}\lambda_{i}^3\,\lambda_{j}&=\sum\limits_{i=1}^{p}\lambda_{i}^3\,\sum\limits_{j=1}^{p}\lambda_{j}-\sum\limits_{j=1}^{p}\lambda_{j}^4=p\textrm{tr}(\bfmath{\rho}^3)-\textrm{tr}(\bfmath{\rho}^4)\,,\\
\sum\limits_{i\neq{j}}\lambda_{i}^2\,\lambda_{j}^2&=\sum\limits_{i=1}^{p}\lambda_{i}^2\sum\limits_{j=1}^{p}\lambda_{j}^2-\sum\limits_{j=1}^{p}\lambda_{j}^4=\left[\textrm{tr}(\bfmath{\rho}^2)\right]^2-\textrm{tr}(\bfmath{\rho}^4)\,,\nonumber\\
\sum\limits_{i\neq{j}\neq{k}}\lambda_{i}^2\,\lambda_{j}\,\lambda_{k}&=\sum\limits_{i=1}^{p}\lambda_{i}^2\sum\limits_{j=1}^{p}\lambda_{j}\sum\limits_{k=1}^{p}\lambda_{k}-\sum\limits_{i\neq{j}}\lambda_{i}^2\,\lambda_{j}^2-2\sum\limits_{i\neq{j}}\lambda_{i}^3\,\lambda_{j}-\sum\limits_{j=1}^{p}\lambda_{j}^4\,,\nonumber\\
&=p^2\,\textrm{tr}(\bfmath{\rho}^2)-\left[\textrm{tr}(\bfmath{\rho}^2)\right]^2-2\,p\,\textrm{tr}(\bfmath{\rho}^3)+2\,\textrm{tr}(\bfmath{\rho}^4)\,,\nonumber\\
\sum_{i\neq{j}\neq{k}\neq{l}}\lambda_{i}\lambda_{j}\lambda_{k}\lambda_{l}&=p^4-6\,\textrm{tr}(\bfmath{\rho}^4)+8\,p\,\textrm{tr}(\bfmath{\rho}^3)+3\left[\textrm{tr}(\bfmath{\rho}^2)\right]^2-6\,p^2\,\textrm{tr}(\bfmath{\rho}^2)\,\,,\nonumber
\end{align}
the 3rd order cumulant $\kappa_3$ of $U=U_i$ can be computed as follow.
\begin{align*}
\kappa_3={\rm E}\left[U^{3}\right]&=\frac{1}{\left[2\,{\rm tr}(\bfmath{\rho}^2)\right]^{\frac{3}{2}}}\,{\rm E}\left[\left(\sum\limits_{j=1}^{p}\lambda_{_j}\xi_{_j}^{2}-p\right)^3\right]\\
&=\frac{1}{\left[2\,\textrm{tr}(\bfmath{\rho}^2)\right]^{\frac{3}{2}}}\,{\rm E}\left[\left(\sum_{j=1}^{p}\lambda_{_j}\xi_{_j}^{2}\right)^3-3\,p\,\left(\sum\limits_{j=1}^{p}\lambda_{_j}\xi_{_j}^{2}\right)^2+3\,p^2\,\sum\limits_{j=1}^{p}\lambda_{_j}\xi_{_j}^{2}-p^3\right]\\
&=\frac{1}{\left[2\,\textrm{tr}(\bfmath{\rho}^2)\right]^{\frac{3}{2}}}\left[A_1+A_2+A_3-p^3\right]
\end{align*}
The term $A_1$ can be calculated as follows 
\begin{align*}
A_1&={\rm E}\left[\left(\sum\limits_{j=1}^{p}\lambda_{j}\,\xi_{_j}^{2}\right)^3\right]\\
&={\rm E}\left[\sum_{j=1}^{p}(\lambda_{j}\xi_{_j}^{2})^3\right]+{\rm E}\left[3\sum\limits_{i\neq{j}}(\lambda_{i}\xi_{_i}^{2})^2\,(\lambda_{j}\xi_{_j}^2)\right]+{\rm E}\left[\sum_{i\neq{j}\neq{k}}(\lambda_{i}\xi_{_i}^{2})(\lambda_{j}\xi_{_j}^2)(\lambda_{k}\xi_{k}^2)\right]\\
&=15\,\sum_{j=1}^{p}\lambda_{j}^3+9\,\sum\limits_{i\neq{j}}\lambda_{i}^2\lambda_{j}+\sum\limits_{i\neq{j}\neq{k}}\lambda_{i}\lambda_{j}\lambda_{k}
=8\,\textrm{tr}(\bfmath{\rho}^3)+6\,p\,\textrm{tr}(\bfmath{\rho}^2)+p^3\,,
\end{align*}
and the terms $A_2$ and $A_3$ are 
\begin{align*}
A_2&=-3\,p\,{\rm E}\left[\left(\sum\limits_{j=1}^{p}\lambda_{j}\xi_{_j}^{2}\right)^2\right]
=-3\,p\,{\rm E}\left[\sum\limits_{j=1}^{p}\lambda_{j}^2\,\xi_{_j}^{4}\right]-3\,p\,{\rm E}\left[\sum\limits_{i\neq{j}}\lambda_{i}\,\lambda_{j}\,\xi_{_i}^{2}\,\xi_{_j}^2\right]
=-6\,p\,\textrm{tr}(\bfmath{\rho}^2)-3\,p^3\,,\\
A_3&=3\,p^2\,{\rm E}\left[\sum\limits_{j=1}^{p}\lambda_{j}\,\xi_{j}^{2}\right]=3\,p^2\,\sum\limits_{j=1}^{p}\lambda_{j}=3\,p^3\,.
\end{align*}
Substituting $A_1$, $A_2$ and $A_3$ back into the equation for $\kappa_{_3}$, we have 

\begin{equation}\kappa_{_3}=\frac{8\,\textrm{tr}\bfmath{\rho}^3}{\left[2\,\textrm{tr}(\bfmath{\rho}^2)\right]^{\frac{3}{2}}}
\end{equation}
To obtain the 2nd order Cornish-Fisher expansion we also require the 4rd order cumulant $\kappa_{_4}$ of $U=U_i$. This can be done in the same manner as above. First notice that the formula for the 4th order cumulant of $U$ reduces to
\begin{align}\label{kappa4comp}
\kappa_{_4}=&{\rm E}\left[U^{4}\right]-3\,{\rm E}\left[U^{2}\right]^2 \nonumber\\
=&\frac{1}{\left[2\,{\rm tr}(\bfmath{\rho}^2)\right]^{2}}\,{\rm E}\left[\sum\limits_{j=1}^{p}\lambda_{j}\,\xi_{j}^{2}-p\right]^4-3\nonumber\\
=&\frac{1}{\left[2\,{\rm tr}(\bfmath{\rho}^2)\right]^{2}}\,{\rm E}\left[\left(\sum\limits_{j=1}^{p}\lambda_{j}\,\xi_{j}^{2}\right)^4-4\,p\,\left(\sum\limits_{j=1}^{p}\lambda_{j}\,\xi_{j}^{2}\right)^3\right.\\
&\left.+6\,p^2\,\left(\sum\limits_{j=1}^{p}\lambda_{j}\,\xi_{j}^{2}\right)^2-4\,p^3\,\sum\limits_{j=1}^{p}\lambda_{j}\,\xi_{j}^{2}+p^4\right]-3\nonumber\\
=&\frac{1}{\left[2\,\textrm{tr}(\bfmath{\rho}^2)\right]^{2}}\,\left[B_1+B_2+B_3+B_4+p^4\right]-3\,.\nonumber
\end{align}
Now, we evaluate the terms $B_1$, $B_2$, $B_3$ and $B_4$ using the equations in  \eqref{Nmoments} and \eqref{sums}. 
\begin{align*}
B_1&={\rm E}\left[\left(\sum\limits_{j=1}^{p}\lambda_{j}\,\xi_{j}^{2}\right)^4\right]\\
&={\rm E}\left[\,\sum\limits_{j=1}^{p}\lambda_{j}^4\,\xi_{j}^{8}+4\,\sum\limits_{i\neq j}\lambda_{i}^3\,\lambda_{j}\xi_{i}^{6}\,\xi_j^2+3\,\sum\limits_{i\neq j}\lambda_{i}^2\,\xi_{i}^{4}\,\lambda_{j}^2\,\xi_{j}^4\right.\\
&\quad \qquad\left.+6\sum_{i\neq{j}\neq{k}}\lambda_{i}^2\,\lambda_{j}\,\lambda_{k}\,\xi_{i}^{4}\,\xi_{j}^2\,\xi_{k}^2+\sum_{i\neq{j}\neq{k}\neq{l}}\lambda_{i}\,\lambda_{j}\,\lambda_{k}\,\lambda_{\ell}\,\xi_{i}^{2}\,\xi_{j}^2\,\xi_{k}^2\,\xi_{\ell}^2\right]\\
&=105\,\sum\limits_{j=1}^{p}\lambda_{j}^4+60\,\sum\limits_{i\neq{j}}\lambda_{i}^3\lambda_{j}+27\,\sum\limits_{i\neq{j}}\lambda_{i}^2\lambda_{j}^2+18\,\sum\limits_{i\neq{j}\neq{k}}\lambda_{i}^2\lambda_{j}\lambda_{k}+\sum\limits_{i\neq{j}\neq{k}\neq{\ell}}\lambda_{i}\lambda_{j}\lambda_{k}\lambda_{\ell}\\
&=48\,{\rm tr}(\bfmath{\rho}^4)+32\,p\,{\rm tr}(\bfmath{\rho}^3) +12\left[{\rm tr}(\bfmath{\rho}^2)\right]^2+12\,p^2\,{\rm tr}(\bfmath{\rho}^2)+p^4\,,\\
B_2&=-4\,p\,{\rm E}\left[\left(\sum\limits_{j=1}^{p}\lambda_{j}\xi_{j}^{2}\right)^{3}\right]\\
&=-4\,p\,{\rm E}\left[\sum\limits_{j=1}^{p}\lambda_{j}^3\,\xi_{j}^{6}+3\,\sum\limits_{i\neq j}\lambda_{i}^2\,\lambda_{j}\,\xi_{i}^{4}\,\xi_{j}^2+\sum\limits_{i\neq j\neq k}\lambda_{i}\,\lambda_{j}\,\lambda_{k}\,\xi_{i}^{2}\,\xi_{j}^2\,\xi_{k}^2\right]\\
&=-4\,p\,\left(15\,\sum\limits_{j=1}^{p}\lambda_{j}^3+9\,\sum\limits_{i\neq j}\,\lambda_{i}^2\,\lambda_{j}+
\sum\limits_{i\neq j \neq k}\lambda_{i}\,\lambda_{j}\,\lambda_{k}\right)\\
&=-32\,p\,{\rm tr}(\bfmath{\rho}^3)-24\,p^2\,{\rm tr}(\bfmath{\rho}^2)-4\,p^4\,,\\
B_3&=6\,p^2\,{\rm E}\left[\left(\sum\limits_{j=1}^{p}\lambda_{j}\,\xi_{j}^{2}\right)^{2}\right]=6\,p^2\,{\rm E}\left[\sum\limits_{j=1}^{p}\lambda_{j}^2\,\xi_{j}^{4}+\sum\limits_{i\neq j}\lambda_{i}\,\lambda_{j}\,\xi_{i}^{2}\,\xi_{j}^2\right]\\
&=18\,p^2\,\sum\limits_{j=1}^{p}\lambda_{j}^2+6\,p^2\,\sum\limits_{i\neq j}\lambda_{i}\,\lambda_{j}
=12\,p^2\,{\rm tr}(\bfmath{\rho}^2)+6\,p^4\,,\\
B_4&=-4\,p^3\,{\rm E}\left[\sum\limits_{j=1}^{p}\lambda_{j}\,\xi_{j}^{2}\right]=-4\,p^3\sum\limits_{j=1}^{p}\lambda_{j}=-4\,p^4\,.
\end{align*}
Substituting $B_1$, $B_2$, $B_3$ and $B_4$ in \eqref{kappa4comp}, we obtain 
\begin{equation}\label{kappa4}
\kappa_{_4}=\frac{48\,{\rm tr}(\bfmath{\rho}^4)+12\left[{\rm tr}(\bfmath{\rho}^2)\right]^2}{\left[2\,{\rm tr}(\bfmath{\rho}^2)\right]^{2}}-3=\frac{12\,{\rm tr}(\bfmath{\rho}^4)}{\left[{\rm tr}(\bfmath{\rho}^2)\right]^{2}}\,\,\cdot
\end{equation}
Now, substituting all required terms in \eqref{C-F}, we obtain the 2nd order Cornish-Fisher for the $(1-\alpha)100\%$ quantile of $U_{_i}$ 
\begin{align*}
\omega_{_{\alpha,\,p}}=z_{_{\alpha}}+\frac{(4\,{\rm tr}(\bfmath{\rho}^{3}))(z_{_{\alpha}}^2-1)}{3\,(2\,{\rm tr}\left(\bfmath{\rho}^{2}\right))^\frac{3}{2}}+\frac{{\rm tr}\left(\bfmath{\rho}^4\right)}{2\,({\rm tr}\left(\bfmath{\rho}^2\right))^{2}}((z_{\alpha}^3-3z_{\alpha})
+\frac{2\,({\rm tr}\left(\bfmath{\rho}^{3}\right))^2}{9\,({\rm tr}\left(\bfmath{\rho}^{2}\right))^{3}}[{5z_{\alpha}-2z_{\alpha}^3}]
\end{align*}

\end {document}